\documentclass{amsart}
\usepackage[foot]{amsaddr}
\usepackage[margin=1in]{geometry}
\usepackage{amsmath, amsfonts, amssymb}
\usepackage{color}
\usepackage{xcolor}
\usepackage{algpseudocode}
\usepackage{algorithm}
\usepackage{siunitx}
\usepackage{graphicx}
\usepackage{arcs}
\usepackage{physics}
\usepackage{bbm}
\usepackage{tikz}
\usepackage{mathtools}
\usepackage{mathrsfs}
\usepackage{caption}
\usepackage{subcaption}
\usepackage{hhline}
\usetikzlibrary{patterns,snakes,arrows.meta}
\colorlet{ColorPink}{black!10}

\usepackage{hyperref}
\usepackage{cleveref}

\newcommand{\bs}{\boldsymbol{s}}
\newcommand{\e}{\mathrm{e}}
\newcommand{\ff}{\mathrm{f}}
\newcommand{\ii}{\mathrm{i}}
\newcommand{\id}{\mathrm{id}}

\newcommand{\sgn}{\mathrm{sgn}}
\newcommand{\dt}{\Delta t}

\newcommand{\highlightyellow}[1]{\colorbox{yellow}{$\displaystyle #1$}}
\newcommand{\highlightgreen}[1]{\colorbox{green}{$\displaystyle #1$}}

\theoremstyle{remark}
\newtheorem{remark}{Remark}

\title[Simulation of open quantum spin chains]{Real-Time Simulation of Open Quantum Spin Chains with Inchworm Method}

\author{Geshuo Wang}
\address[Geshuo Wang]{Department of Mathematics, National University of Singapore,
  Level 4, Block S17, 10 Lower Kent Ridge Road, Singapore 119076}
\email{geshuowang@u.nus.edu}

\author{Zhenning Cai}
\address[Zhenning Cai]{Department of Mathematics, National University of Singapore,
  Level 4, Block S17, 10 Lower Kent Ridge Road, Singapore 119076}
\email{matcz@nus.edu.sg}

\thanks{Zhenning Cai's work was supported by the Academic Research Fund of the Ministry of Education of Singapore under grant A-8000965-00-00.}


\begin{document}

\maketitle

\begin{abstract}
We study the real-time simulation of open quantum systems,
 where the system is modeled by a spin chain,
 with each spin associated with its own harmonic bath.
Our method couples the inchworm method for the spin-boson model and the modular path integral methodology for spin systems.
In particular, the introduction of the inchworm method can significantly suppress the numerical sign problem.
Both methods are tweaked to make them work seamlessly with each other.
We represent our approach in the language of diagrammatic methods,
 and analyze the asymptotic behavior of the computational cost.
Extensive numerical experiments are done to validate our method.
\end{abstract}

\section{Introduction}
An open quantum system refers to a quantum-mechanical system coupled to an environment.
The coupling can significantly affect the quantum dynamics,
 resulting in effects such as quantum dissipation and quantum decoherence.
It can also lead to non-Markovian evolution of the quantum system,
 posing significant challenges in the numerical simulation.
Nevertheless, the study of open quantum systems is becoming increasingly important and has practical applications in many fields \cite{nielsen2002quantum,breuer2002theory,carmichael2009open}, as real-world systems are never completely isolated.

In the simulation of open quantum systems, a simple harmonic bath is generally assumed so that the effect of the bath on the system can be analytically given by the bath influence functional \cite{feynman1963theory},
 allowing the path integral approach \cite{feynman1948space} to be used to formulate the system dynamics.
One classical method based on path integrals is the quasi-adiabatic propagator path integral (QuAPI) \cite{makri1992improved}.
Other methods have been developed based on QuAPI to improve simulation efficiency by reducing computational complexity or enhancing computational accuracy, including the iterative QuAPI method \cite{makri1995numerical,makri1998quantum},
the blip decomposition of the path integral \cite{makri2014blip,makri2017blip}
and differential equation-based path integral method (DEBPI)
\cite{wang2022differential}.
Due to the non-Markovian nature of the dynamics,
 the path-integral-based methods often suffer from increasing memory costs for longer simulation time.
The small matrix decomposition of the path integral (SMatPI) \cite{makri2020small1,makri2020small2,makri2021small1,makri2021small2}, however, has successfully overcome the problem by summarizing the contribution of the paths into small matrices representing the kernel of the quantum master equation.

An alternative approach to dealing with the high memory cost in simulating quantum systems is to use the quantum Monte Carlo method to evaluate the high-dimensional integrals in the Dyson series \cite{dyson1949radiation}.
However, the Monte Carlo method introduces stochastic errors and can lead to the so-called ``sign problem'' for highly oscillatory integrands \cite{loh1990sign}.
To relieve the sign problem, the inchworm Monte Carlo method was developed in \cite{chen2017inchworm1,chen2017inchworm2},
 which takes the idea of bold diagrammatic Monte Carlo method introduced in \cite{prokof2007bold}.
The idea is to compute quantum propagators for shorter time intervals,
 and then combine them into the propagators of longer time intervals.
The extension of the propagators can also be formulated into an integro-differential equation \cite{cai2020inchworm},
 so that classical numerical methods can be applied.
The inchworm Monte Carlo method has been proven to be successful in reducing the severity of sign problem \cite{dong2017quantum,ridley2018numerically,eidelstein2020multiorbital,cai2022numerical}.
Some efficient numerical methods for solving the integro-differential equation has been discussed in \cite{cai2022fast,cai2022bold}.

The methods discussed above are mainly focused on simple systems such as a single spin or other systems with a small number of possible states,
 since the dimension of the Hilbert space for a system grows exponentially with the number of particles. 
As a result, simulating more complex systems requires new approaches. 
One such approach is the method of modular path integral (MPI) \cite{makri2018modular,makri2018communication,kundu2019modular,kundu2020modular},
which leads to linear scaling with the number of particles.
Other methods apply tensor train decomposition to keep the memory cost low for large systems \cite{white1992density,schollwock2011density,orus2014practical,ren2018time}, which utilizes low-rank approximations to reduce the computational and memory cost.
In these methods, a typical system under consideration is the Ising chain model, a one-dimensional chain of interacting spins \cite{ising1924beitrag,deGennes1963collective}.
The Ising model has wide application in magnetism \cite{luck1993critical}, neuroscience \cite{hopfield1982neural,schneidman2006weak} and many other fields.
The dynamics of closed Ising chains is well-studied in the literature \cite{glauber1963time,pfeuty1970one,droz1986critical,luscombe1987nonuniversal,fisher1992random,dziarmaga2005dynamics}.
Recently, there has been more research focusing on the dissipative Ising chain \cite{werner2005phase,hoyos2012dissipation,takada2016critical,schollwock2011density}. 

This paper focuses on the evolution of an Ising chain coupled with harmonic baths,
 which are characterized by the Ohmic spectral density \cite{caldeira1983path}.
The Ising model used in this study is introduced in \Cref{sec_Ising_chain}. 
In \Cref{sec_diagram_represent}, 
we propose a diagrammatic representation of the model based on the special structure of the Ising chain.
The computation of the diagrams is introduced in detail in \Cref{sec_inchworm} and \Cref{sec_resum}.
\Cref{sec_inchworm} mainly discusses the computation of diagrams for each single spin,
 and \Cref{sec_resum} contains the algorithm for merging the diagrams.
The estimation of the computational cost is given in \Cref{sec_computational_cost}, and numerical experiments are given in \Cref{sec_numerical}.
Finally, in \Cref{sec_conclusion}, we provide some concluding remarks and introduce possible future works inspired by our results.
\section{Ising Chain with Spin-Bath Coupling}
\label{sec_Ising_chain}
\label{sec_model}
This section provides a brief introduction to the model studied in this paper, 
which is an Ising chain coupled with baths consisting of harmonic oscillators.
In this model, the baths for different spins are not directly coupled.

An isolated Ising chain is a chain of spins in which each spin couples with its nearest neighbors \cite{ising1924beitrag}.
The Hamiltonian for an Ising chain with $K$ spins is generally given by
\begin{equation*}
    H_{\mathrm{Ising}}
    = \sum_{k=1}^K H_s^{(k)}
    + \sum_{k=1}^{K-1} U^{(k)} \otimes V^{(k+1)}.
\end{equation*}
where
\begin{equation*}
    H_s^{(k)} = \epsilon^{(k)} \sigma_z^{(k)} + \Delta^{(k)} \sigma_x^{(k)}
\end{equation*}
with $\sigma_x^{(k)},\sigma_z^{(k)}$ being Pauli matrices
 for the $k$th spin in the chain.
The parameter $\epsilon^{(k)}$ describes the energy difference between two spin states 
and $\Delta^{(k)}$ is the frequency of the spin flipping.
The term $U^{(k)}\otimes V^{(k+1)}$ describes the nearest-neighbor coupling between the $k$th and $(k+1)$th spins.

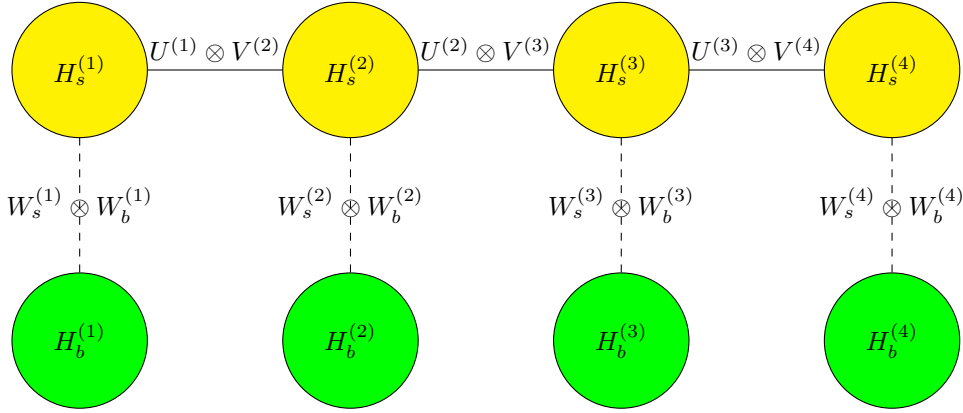
\begin{figure}
\centering
\begin{tikzpicture}[scale=0.9]
\filldraw[fill=yellow] (0,0) circle (1cm);
\node[text=black] at (0,0) {$H_s^{(1)}$};
\filldraw[fill=yellow] (4,0) circle (1cm);
\node[text=black] at (4,0) {$H_s^{(2)}$};
\filldraw[fill=yellow] (8,0) circle (1cm);
\node[text=black] at (8,0) {$H_s^{(3)}$};
\filldraw[fill=yellow] (12,0) circle (1cm);
\node[text=black] at (12,0) {$H_s^{(4)}$};
\draw[-] (1,0) -- (3,0);
\draw[-] (5,0) -- (7,0);
\draw[-] (9,0) -- (11,0);
\node[text=black] at (2,0.3) {$U^{(1)}\otimes V^{(2)}$};
\node[text=black] at (6,0.3) {$U^{(2)}\otimes V^{(3)}$};
\node[text=black] at (10,0.3) {$U^{(3)}\otimes V^{(4)}$};
\filldraw[fill=green] (0,-4) circle (1cm);
\node[text=black] at (0,-4) {$H_b^{(1)}$};
\filldraw[fill=green] (4,-4) circle (1cm);
\node[text=black] at (4,-4) {$H_b^{(2)}$};
\filldraw[fill=green] (8,-4) circle (1cm);
\node[text=black] at (8,-4) {$H_b^{(3)}$};
\filldraw[fill=green] (12,-4) circle (1cm);
\node[text=black] at (12,-4) {$H_b^{(4)}$};
\draw[dashed] (0,-1) -- (0,-3);
\draw[dashed] (4,-1) -- (4,-3);
\draw[dashed] (8,-1) -- (8,-3);
\draw[dashed] (12,-1) -- (12,-3);
\node[text=black] at (0,-2) {$W_s^{(1)}\otimes W_b^{(1)}$};
\node[text=black] at (4,-2) {$W_s^{(2)}\otimes W_b^{(2)}$};
\node[text=black] at (8,-2) {$W_s^{(3)}\otimes W_b^{(3)}$};
\node[text=black] at (12,-2) {$W_s^{(4)}\otimes W_b^{(4)}$};
\end{tikzpicture}
\caption{Illustration of the Ising chain with four spins where each spin is coupled with a harmonic bath. Each yellow circle stands for a spin and each green circle stands for a harmonic bath. A solid line between two spins indicate the coupling given by $U^{(k)}\otimes V^{(k+1)}$ and a dashed line between a yellow circle and a green circle stands for the coupling relation between a spin and its bath.}
\label{diagram_model}
\end{figure}

In this paper, a more complicated case is studied
 where each spin in the Ising chain is coupled with a harmonic bath.
The total Hamiltonian for the whole system-bath is then given by
\begin{equation}
\label{total_Hamiltonian}
    H = H_{\mathrm{Ising}} + \sum_{k=1}^K H_b^{(k)} + \sum_{k=1}^K W_s^{(k)} \otimes W_b^{(k)}
\end{equation}
where
\begin{equation*}
    H_b^{(k)} = \sum_{j} \frac{1}{2} \left[\left(\hat{p}_{j}^{(k)}\right)^2 + \left(\omega_{j}^{(k)}\right)^2 \left(\hat{q}_{j}^{(k)}\right)^2\right],
    \quad W_s^{(k)} = \sigma_z^{(k)},
    \quad W_b^{(k)} = \sum_{j} c_{j}^{(k)} \hat{q}_{j}^{(k)}.
\end{equation*}
In this expression, $\hat{p}_j^{(k)}$ and $\hat{q}_j^{(k)}$ are the momentum operator and the position operator of the $j$th harmonic oscillator in the bath of the $k$th spin, respectively.
$\omega_j^{(k)}$ is the frequency of the $j$th harmonic oscillator in the bath of the $k$th spin and $c_j^{(k)}$ is the coupling intensity between the $k$th spin and the $j$th oscillator in its bath.
\Cref{diagram_model} illustrates the overall Hamiltonian and the coupling relation in this model more intuitively
 with a Ising chain with 4 spins.
Similar to the assumption in \cite{makri2018modular}, 
 in the paper, the baths for different spins are not directly coupled with each other.

Similar to \cite{makri2018communication,makri2018modular}, we simply use $U^{(k)} = V^{(k)}$ so that our method can be better illustrated by diagrams in the following sections.
The method discussed in this paper is also applicable to a more general system $U^{(k)} \not= V^{(k)}$. 

As for the initial condition, the spins and the baths are assumed to be decoupled.
More specifically, 
 the $k$th spin is assumed to be in the state $\ket{\varsigma^{(k)}}$
 and the baths are at their thermal equilibriums.
The initial density matrix for the whole system is then given by
\begin{equation}
\label{eq_Initial_condition_general}
    \rho(0)
    = \bigotimes_{k=1}^K \rho^{(k)} (0)
    = \bigotimes_{k=1}^K \left( \rho_s^{(k)}(0) \otimes \rho_b^{(k)}(0) \right)
    = \bigotimes_{k=1}^K \left( \dyad{\varsigma^{(k)}} \otimes
    \frac{\exp(-\beta^{(k)} H_b^{(k)})}{\tr\left(\exp(-\beta^{(k)} H_b^{(k)})\right)}
    \right)
\end{equation}
 where $\beta^{(k)}$ is the inverse temperature for the $k$th bath \cite{makri1998quantum}.

\section{Diagrammatic Representation of the Path Integral}
\label{sec_diagram_represent}
In this section, we rewrite the evolution of the spin chain system using path integrals, so that the computation of each spin can be decoupled. Such an approach has been studied in many previous works \cite{makri2018modular, cai2018quantum}, and here we are going to represent the path integrals using diagrams to facilitate our future discussions.

We first split the total Hamiltonian in \cref{total_Hamiltonian} into two parts $H = H_0 + V$, where
\begin{equation}
\label{def_H0_V}
    H_0 \coloneqq \sum_{k=1}^K H_0^{(k)} 
    \coloneqq \sum_{k=1}^K
    \left(H_s^{(k)} + H_b^{(k)} + W_s^{(k)} \otimes W_b^{(k)}\right),\quad
    V \coloneqq \sum_{k=1}^{K-1} V^{(k)}\otimes V^{(k+1)}.
\end{equation}
Below, we will assume that the interaction between spins $V$ is a perturbation of the unperturbed Hamiltonian $H_0$,
 and describe the dynamics in the interaction picture.
Given an observable $O = O_s \otimes \mathrm{Id}_b$,
 we can define the following propagator
\begin{displaymath}
    G(-t,t) = \e^{-\ii H_0 t} \e^{\ii H t} O \e^{-\ii H t}\e^{\ii H_0 t},
\end{displaymath}
which can be expanded into the following Dyson series 
\begin{equation}
    \label{propagator_multiple_spins}
     G(-t,t) = \sum_{N=0}^\infty
     \int_{-t\leqslant \bs\leqslant t}
     \left(\prod_{n=1}^N \ii\, \sgn(s_n)\right)
     \mathcal{T}[V_I(s_N) \cdots V_I(s_1) O_{s,I}(0)] \dd\bs
\end{equation}
where 
\begin{equation*}
    V_I(s_n) \coloneqq \e^{-\ii H_0 \vert s_n \vert} V \e^{\ii H_0 \vert s_n \vert}, \quad
    O_{s,I}(0) = O_s
\end{equation*}
and $\mathcal{T}$ is the time ordering operator
that sorts all the operators in the time descending order. 
The integrals in the equation is interpreted as
\begin{equation*}
    \int_{-t\leqslant \boldsymbol{s}\leqslant t} 
    \text{(integrand)}
    \dd\boldsymbol{s}
    = \int_{-t}^{t} 
    \int_{-t}^{s_N}
    \dots
    \int_{-t}^{s_2}
    \text{(integrand)}
    \dd s_1 
    \dots
    \dd s_{N-1}
    \dd s_N
\end{equation*}
Note that the coefficient $\prod_{n=1}^N \ii\, \sgn(s_n)$ comes from the coupling operators $V$, meaning that each $V_I(s_n)$ is attached by $\ii$ or $-\ii$ according to the sign of $s_n$.
With this propagator, the expectation of the observable can be expressed by \cite{cai2020inchworm} 
\begin{equation}
    \label{eq_observable}
    \expval{O_s(t)} 
    = \tr(\rho_I(t) G(-t,t))
\end{equation}
with $\rho_I(t) = \e^{-\ii H_0 t} \rho(0) \e^{\ii H_0 t}$.
If the observable has the form $O_s = O_s^{(1)} \otimes \dots \otimes O_s^{(K)}$,
 we can plug the definition of $V$ in \cref{def_H0_V} into the Dyson series \cref{propagator_multiple_spins}, so that the integrand will show $N$ summation symbols, and each summand can be written in the tensor product form.
Precisely speaking, for the $k$th spin, the summand has the form:
\begin{equation}
\label{Gcal_def}
    \mathcal{G}^{(k)}(\boldsymbol{s}')
    = \left(\prod_{n'=1}^{N'} \sqrt{\ii\,\sgn(s_{n'}')} \right)
    \mathcal{T}\left[
    V_I^{(k)}(s_{N'}') \dots V_I^{(k)}(s_1') O_{s,I}^{(k)}(0)
    \right],
\end{equation}
where $\boldsymbol{s}'$ is a subsequence of $\boldsymbol{s}$ of length $N' \leqslant N$.
In particular, if $\boldsymbol{s'}$ is an empty sequence, we use the notation $\mathcal{G}^{(k)}(\varnothing)\coloneqq O_{s,I}^{(k)}(0)$ to denote the above quantity.
Here we have again used the interaction picture: 
\begin{equation*}
    V_I^{(k)}(s_n) \coloneqq
    \e^{- \ii H_0^{(k)} \vert s_n\vert}
    V^{(k)}
    \e^{\ii H_0^{(k)} \vert s_n \vert},
    \quad
    O_{s,I}^{(k)}(0) = O_s^{(k)}.
\end{equation*}

In \cref{Gcal_def},
 the subsequence $\bs'$ depends on the number of operators $V^{(k)}$ appearing in the summand,
 and the reason for the square root is that the term $\ii V^{(k)} \otimes V^{(k+1)}$ or $-\ii V^{(k)} \otimes V^{(k+1)}$,
 appearing in the expansion of $\ii V$ or $-\ii V$,
 is separated into the terms $\mathcal{G}^{(k)}$ and $\mathcal{G}^{(k+1)}$ after decomposition.
In this work, we stick to the choice $\sqrt{\ii} = \e^{\ii\pi/4}$ and $\sqrt{-\ii} = \e^{-\ii \pi /4}$.

With these propagators, the terms in \cref{propagator_multiple_spins}
can be represented by the sum of integrals whose integrands are tensor products of $\mathcal{G}^{(k)}(\boldsymbol{s})$.
For example, when $N=1$ and $K=4$,
we have
\begin{equation} 
\label{eq:m1n4}
\begin{split}
    &\int_{-t}^t
    \ii \, \sgn(s_1)
    \mathcal{T}\left[V_I(s_1)O_{s,I}(0)\right]
    \dd s_1
    =\int_{-t}^t \mathcal{G}^{(1)}(s_1)
     \otimes \mathcal{G}^{(2)}(s_1)
     \otimes \mathcal{G}^{(3)}(\varnothing)
     \otimes \mathcal{G}^{(4)}(\varnothing) \dd s_1\\
    &\hspace{1cm}+\int_{-t}^t \mathcal{G}^{(1)}(\varnothing)
     \otimes \mathcal{G}^{(2)}(s_1)
     \otimes \mathcal{G}^{(3)}(s_1)
     \otimes \mathcal{G}^{(4)}(\varnothing) \dd s_1
    +\int_{-t}^t \mathcal{G}^{(1)}(\varnothing)
     \otimes \mathcal{G}^{(2)}(\varnothing)
     \otimes \mathcal{G}^{(3)}(s_1)
     \otimes \mathcal{G}^{(4)}(s_1) \dd s_1.
\end{split}
\end{equation}
In this equation,
 different spins are separated inside the integrals, allowing us to perform computations for each spin independently.
For simplicity, we may express the above equation as a diagrammatic equation:
\begin{equation}
\begin{tikzpicture}[baseline=0]
\filldraw[fill=black] (-1,-0.1) rectangle (1,0.1);
\node[text=black,anchor=north] at (-1,0) {$-t$};
\node[text=black,anchor=north] at (1,0) {$t$};
\draw plot[only marks, mark=x, mark options={color=red, scale=2, ultra thick}] coordinates {(-0.5,0)};
\node[text=black,anchor=north] at (-0.5,0) {$s_1$};
\end{tikzpicture}
= 
\begin{tikzpicture}[baseline=0]
\filldraw[fill=lightgray] (-1,0.55) rectangle (1,0.65);
\filldraw[fill=lightgray] (-1,0.15) rectangle (1,0.25);
\filldraw[fill=lightgray] (-1,-0.25) rectangle (1,-0.15);
\filldraw[fill=lightgray] (-1,-0.65) rectangle (1,-0.55);
\node[text=red] at (-0.5,0.6) {$\times$};
\node[text=red] at (-0.5,0.2) {$\times$};
\draw[black, densely dotted, line width = 1pt] (-0.5,0.6) -- (-0.5,0.2);
\node[text=black,anchor=north] at (-1,-0.6) {$-t$};
\node[text=black,anchor=north] at (1,-0.6) {$t$};
\node[text=black,anchor=north] at (-0.5,-0.6) {$s_1$};
\end{tikzpicture}
+
\begin{tikzpicture}[baseline=0]
\filldraw[fill=lightgray] (-1,0.55) rectangle (1,0.65);
\filldraw[fill=lightgray] (-1,0.15) rectangle (1,0.25);
\filldraw[fill=lightgray] (-1,-0.25) rectangle (1,-0.15);
\filldraw[fill=lightgray] (-1,-0.65) rectangle (1,-0.55);
\node[text=red] at (-0.5,0.2) {$\times$};
\node[text=red] at (-0.5,-0.2) {$\times$};
\draw[black, densely dotted, line width = 1pt] (-0.5,0.2) -- (-0.5,-0.2);
\node[text=black,anchor=north] at (-1,-0.6) {$-t$};
\node[text=black,anchor=north] at (1,-0.6) {$t$};
\node[text=black,anchor=north] at (-0.5,-0.6) {$s_1$};
\end{tikzpicture}
+
\begin{tikzpicture}[baseline=0]
\filldraw[fill=lightgray] (-1,0.55) rectangle (1,0.65);
\filldraw[fill=lightgray] (-1,0.15) rectangle (1,0.25);
\filldraw[fill=lightgray] (-1,-0.25) rectangle (1,-0.15);
\filldraw[fill=lightgray] (-1,-0.65) rectangle (1,-0.55);
\node[text=red] at (-0.5,-0.2) {$\times$};
\node[text=red] at (-0.5,-0.6) {$\times$};
\draw[black, densely dotted, line width = 1pt] (-0.5,-0.6) -- (-0.5,-0.2);
\node[text=black,anchor=north] at (-1,-0.6) {$-t$};
\node[text=black,anchor=north] at (1,-0.6) {$t$};
\node[text=black,anchor=north] at (-0.5,-0.6) {$s_1$};
\end{tikzpicture}
\end{equation}%
In this diagrammatic equation,
 the bold line on the left-hand side represents an operator acting on all spins.
The red cross indicates that only one coupling operator at time $s_1$ exists in the integral. On the right-hand side,
 each gray line represents a single spin.
Since each interaction operator $V$ consists of three terms, each acting on two neighboring spins,
 we have three diagrams on the right-hand side,
 and each diagram includes two red crosses connected by a dotted line, 
 indicating the two involved spins.
By comparison with \eqref{eq:m1n4},
 we can find that every diagram on the right-hand side is an integral with respect to $s_1$,
 and the $k$th line corresponds to the expression $\mathcal{G}^{(k)}(\dots)$,
 where the ellipses should be filled with the time points of the red crosses. In this case, the ellipses can only be a single point $s_1$ or an empty set.

Similarly, for the term with two coupling operators ($N=2$), the expansion is
\begin{equation}
\begin{split}
\begin{tikzpicture}[baseline=0]
\filldraw[fill=black] (-1,-0.1) rectangle (1,0.1);
\node[text=black,anchor=north] at (-1,0) {$-t$};
\node[text=black,anchor=north] at (1,0) {$t$};
\draw plot[only marks, mark=x, mark options={color=red, scale=2, ultra thick}] coordinates {(-0.5,0)};
\node[text=black,anchor=north] at (-0.5,0) {$s_1$};
\draw plot[only marks, mark=x, mark options={color=red, scale=2, ultra thick}] coordinates {(0.3,0)};
\node[text=black,anchor=north] at (-0.5,0) {$s_1$};
\node[text=black,anchor=north] at (0.3,0) {$s_2$};
\end{tikzpicture}
&= 
\begin{tikzpicture}[baseline=0]
\filldraw[fill=lightgray] (-1,0.55) rectangle (1,0.65);
\filldraw[fill=lightgray] (-1,0.15) rectangle (1,0.25);
\filldraw[fill=lightgray] (-1,-0.25) rectangle (1,-0.15);
\filldraw[fill=lightgray] (-1,-0.65) rectangle (1,-0.55);
\node[text=red] at (-0.5,0.6) {$\times$};
\node[text=red] at (-0.5,0.2) {$\times$};
\draw[black, densely dotted, line width = 1pt] (-0.5,0.6) -- (-0.5,0.2);
\node[text=red] at (0.3,0.6) {$\times$};
\node[text=red] at (0.3,0.2) {$\times$};
\draw[black, densely dotted, line width = 1pt] (0.3,0.6) -- (0.3,0.2);
\node[text=black,anchor=north] at (-1,-0.6) {$-t$};
\node[text=black,anchor=north] at (1,-0.6) {$t$};
\node[text=black,anchor=north] at (-0.5,-0.6) {$s_1$};
\node[text=black,anchor=north] at (0.3,-0.6) {$s_2$};
\end{tikzpicture}
+
\begin{tikzpicture}[baseline=0]
\filldraw[fill=lightgray] (-1,0.55) rectangle (1,0.65);
\filldraw[fill=lightgray] (-1,0.15) rectangle (1,0.25);
\filldraw[fill=lightgray] (-1,-0.25) rectangle (1,-0.15);
\filldraw[fill=lightgray] (-1,-0.65) rectangle (1,-0.55);
\node[text=red] at (-0.5,0.6) {$\times$};
\node[text=red] at (-0.5,0.2) {$\times$};
\draw[black, densely dotted, line width = 1pt] (-0.5,0.6) -- (-0.5,0.2);
\node[text=red] at (0.3,0.2) {$\times$};
\node[text=red] at (0.3,-0.2) {$\times$};
\draw[black, densely dotted, line width = 1pt] (0.3,0.2) -- (0.3,-0.2);
\node[text=black,anchor=north] at (-1,-0.6) {$-t$};
\node[text=black,anchor=north] at (1,-0.6) {$t$};
\node[text=black,anchor=north] at (-0.5,-0.6) {$s_1$};
\node[text=black,anchor=north] at (0.3,-0.6) {$s_2$};
\end{tikzpicture}
+
\begin{tikzpicture}[baseline=0]
\filldraw[fill=lightgray] (-1,0.55) rectangle (1,0.65);
\filldraw[fill=lightgray] (-1,0.15) rectangle (1,0.25);
\filldraw[fill=lightgray] (-1,-0.25) rectangle (1,-0.15);
\filldraw[fill=lightgray] (-1,-0.65) rectangle (1,-0.55);
\node[text=red] at (-0.5,0.6) {$\times$};
\node[text=red] at (-0.5,0.2) {$\times$};
\draw[black, densely dotted, line width = 1pt] (-0.5,0.6) -- (-0.5,0.2);
\node[text=red] at (0.3,-0.2) {$\times$};
\node[text=red] at (0.3,-0.6) {$\times$};
\draw[black, densely dotted, line width = 1pt] (0.3,-0.2) -- (0.3,-0.6);
\node[text=black,anchor=north] at (-1,-0.6) {$-t$};
\node[text=black,anchor=north] at (1,-0.6) {$t$};
\node[text=black,anchor=north] at (-0.5,-0.6) {$s_1$};
\node[text=black,anchor=north] at (0.3,-0.6) {$s_2$};
\end{tikzpicture} \\
&+
\begin{tikzpicture}[baseline=0]
\filldraw[fill=lightgray] (-1,0.55) rectangle (1,0.65);
\filldraw[fill=lightgray] (-1,0.15) rectangle (1,0.25);
\filldraw[fill=lightgray] (-1,-0.25) rectangle (1,-0.15);
\filldraw[fill=lightgray] (-1,-0.65) rectangle (1,-0.55);
\node[text=red] at (-0.5,0.2) {$\times$};
\node[text=red] at (-0.5,-0.2) {$\times$};
\draw[black, densely dotted, line width = 1pt] (-0.5,0.2) -- (-0.5,-0.2);
\node[text=red] at (0.3,0.6) {$\times$};
\node[text=red] at (0.3,0.2) {$\times$};
\draw[black, densely dotted, line width = 1pt] (0.3,0.6) -- (0.3,0.2);
\node[text=black,anchor=north] at (-1,-0.6) {$-t$};
\node[text=black,anchor=north] at (1,-0.6) {$t$};
\node[text=black,anchor=north] at (-0.5,-0.6) {$s_1$};
\node[text=black,anchor=north] at (0.3,-0.6) {$s_2$};
\end{tikzpicture}
+
\begin{tikzpicture}[baseline=0]
\filldraw[fill=lightgray] (-1,0.55) rectangle (1,0.65);
\filldraw[fill=lightgray] (-1,0.15) rectangle (1,0.25);
\filldraw[fill=lightgray] (-1,-0.25) rectangle (1,-0.15);
\filldraw[fill=lightgray] (-1,-0.65) rectangle (1,-0.55);
\node[text=red] at (-0.5,0.2) {$\times$};
\node[text=red] at (-0.5,-0.2) {$\times$};
\draw[black, densely dotted, line width = 1pt] (-0.5,0.2) -- (-0.5,-0.2);
\node[text=red] at (0.3,0.2) {$\times$};
\node[text=red] at (0.3,-0.2) {$\times$};
\draw[black, densely dotted, line width = 1pt] (0.3,0.2) -- (0.3,-0.2);
\node[text=black,anchor=north] at (-1,-0.6) {$-t$};
\node[text=black,anchor=north] at (1,-0.6) {$t$};
\node[text=black,anchor=north] at (-0.5,-0.6) {$s_1$};
\node[text=black,anchor=north] at (0.3,-0.6) {$s_2$};
\end{tikzpicture}
+
\begin{tikzpicture}[baseline=0]
\filldraw[fill=lightgray] (-1,0.55) rectangle (1,0.65);
\filldraw[fill=lightgray] (-1,0.15) rectangle (1,0.25);
\filldraw[fill=lightgray] (-1,-0.25) rectangle (1,-0.15);
\filldraw[fill=lightgray] (-1,-0.65) rectangle (1,-0.55);
\node[text=red] at (-0.5,0.2) {$\times$};
\node[text=red] at (-0.5,-0.2) {$\times$};
\draw[black, densely dotted, line width = 1pt] (-0.5,0.2) -- (-0.5,-0.2);
\node[text=red] at (0.3,-0.2) {$\times$};
\node[text=red] at (0.3,-0.6) {$\times$};
\draw[black, densely dotted, line width = 1pt] (0.3,-0.2) -- (0.3,-0.6);
\node[text=black,anchor=north] at (-1,-0.6) {$-t$};
\node[text=black,anchor=north] at (1,-0.6) {$t$};
\node[text=black,anchor=north] at (-0.5,-0.6) {$s_1$};
\node[text=black,anchor=north] at (0.3,-0.6) {$s_2$};
\end{tikzpicture} \\
&+
\begin{tikzpicture}[baseline=0]
\filldraw[fill=lightgray] (-1,0.55) rectangle (1,0.65);
\filldraw[fill=lightgray] (-1,0.15) rectangle (1,0.25);
\filldraw[fill=lightgray] (-1,-0.25) rectangle (1,-0.15);
\filldraw[fill=lightgray] (-1,-0.65) rectangle (1,-0.55);
\node[text=red] at (-0.5,-0.2) {$\times$};
\node[text=red] at (-0.5,-0.6) {$\times$};
\draw[black, densely dotted, line width = 1pt] (-0.5,-0.2) -- (-0.5,-0.6);
\node[text=red] at (0.3,0.6) {$\times$};
\node[text=red] at (0.3,0.2) {$\times$};
\draw[black, densely dotted, line width = 1pt] (0.3,0.6) -- (0.3,0.2);
\node[text=black,anchor=north] at (-1,-0.6) {$-t$};
\node[text=black,anchor=north] at (1,-0.6) {$t$};
\node[text=black,anchor=north] at (-0.5,-0.6) {$s_1$};
\node[text=black,anchor=north] at (0.3,-0.6) {$s_2$};
\end{tikzpicture}
+
\begin{tikzpicture}[baseline=0]
\filldraw[fill=lightgray] (-1,0.55) rectangle (1,0.65);
\filldraw[fill=lightgray] (-1,0.15) rectangle (1,0.25);
\filldraw[fill=lightgray] (-1,-0.25) rectangle (1,-0.15);
\filldraw[fill=lightgray] (-1,-0.65) rectangle (1,-0.55);
\node[text=red] at (-0.5,-0.2) {$\times$};
\node[text=red] at (-0.5,-0.6) {$\times$};
\draw[black, densely dotted, line width = 1pt] (-0.5,-0.2) -- (-0.5,-0.6);
\node[text=red] at (0.3,0.2) {$\times$};
\node[text=red] at (0.3,-0.2) {$\times$};
\draw[black, densely dotted, line width = 1pt] (0.3,0.2) -- (0.3,-0.2);
\node[text=black,anchor=north] at (-1,-0.6) {$-t$};
\node[text=black,anchor=north] at (1,-0.6) {$t$};
\node[text=black,anchor=north] at (-0.5,-0.6) {$s_1$};
\node[text=black,anchor=north] at (0.3,-0.6) {$s_2$};
\end{tikzpicture}
+
\begin{tikzpicture}[baseline=0]
\filldraw[fill=lightgray] (-1,0.55) rectangle (1,0.65);
\filldraw[fill=lightgray] (-1,0.15) rectangle (1,0.25);
\filldraw[fill=lightgray] (-1,-0.25) rectangle (1,-0.15);
\filldraw[fill=lightgray] (-1,-0.65) rectangle (1,-0.55);
\node[text=red] at (-0.5,-0.2) {$\times$};
\node[text=red] at (-0.5,-0.6) {$\times$};
\draw[black, densely dotted, line width = 1pt] (-0.5,-0.2) -- (-0.5,-0.6);
\node[text=red] at (0.3,-0.2) {$\times$};
\node[text=red] at (0.3,-0.6) {$\times$};
\draw[black, densely dotted, line width = 1pt] (0.3,-0.2) -- (0.3,-0.6);
\node[text=black,anchor=north] at (-1,-0.6) {$-t$};
\node[text=black,anchor=north] at (1,-0.6) {$t$};
\node[text=black,anchor=north] at (-0.5,-0.6) {$s_1$};
\node[text=black,anchor=north] at (0.3,-0.6) {$s_2$};
\end{tikzpicture}.
\end{split}
\end{equation}%
Here the left-hand side corresponds to the two-dimensional integral in  \cref{propagator_multiple_spins}.
On the right-hand side,
 we have nine diagrams since both interaction operators at $s_1$ and $s_2$ have three choices.
For general $N$ and $K$,
 the number of diagrams should be $(K-1)^N$.
In particular, 
 for the first term in \cref{propagator_multiple_spins}
 where no interaction exists,
no integral is required and we have
\begin{equation*}
    O_s(0) = O_s = O_s^{(1)} \otimes O_s^{(2)} \otimes O_s^{(3)} \otimes O_s^{(4)} \otimes
    = \mathcal{G}^{(1)} (\varnothing)
    \otimes \mathcal{G}^{(2)} (\varnothing)
    \otimes \mathcal{G}^{(3)} (\varnothing)
    \otimes \mathcal{G}^{(4)} (\varnothing),
\end{equation*}
 which can be represented by the following diagrammatic equation:
\begin{equation*}
\begin{tikzpicture}[baseline=0]
\filldraw[fill=black] (-1,-0.1) rectangle (1,0.1);
\node[text=black,anchor=north] at (-1,0) {$-t$};
\node[text=black,anchor=north] at (1,0) {$t$};
\end{tikzpicture}
= 
\begin{tikzpicture}[baseline=0]
\filldraw[fill=lightgray] (-1,0.55) rectangle (1,0.65);
\filldraw[fill=lightgray] (-1,0.15) rectangle (1,0.25);
\filldraw[fill=lightgray] (-1,-0.25) rectangle (1,-0.15);
\filldraw[fill=lightgray] (-1,-0.65) rectangle (1,-0.55);
\node[text=black,anchor=north] at (-1,-0.6) {$-t$};
\node[text=black,anchor=north] at (1,-0.6) {$t$};
\end{tikzpicture}.
\end{equation*}
As a result, the final diagrammatic expansion of 
$G(-t,t)$ is 
\input{diagram_equation_full_propagator.tex}%
where the right-hand side includes all possible connections between neighboring spins.

The advantage of this expansion is two-fold:
\begin{enumerate}
  \item For each diagram, when the time points $s_1, \cdots, s_N$ are fixed, the $k$th line with crosses is mathematically represented by $\mathcal{G}^{(k)}(\bs)$, which involves only one spin, so that it can be computed relatively easily.
  \item We can shuffle the diagrams and truncate the series appropriately to obtain efficient algorithms.
\end{enumerate}
The idea for the computation of each line on the right-hand side will be based on an efficient path integral method known as the inchworm method \cite{chen2017inchworm1, chen2017inchworm2, cai2020inchworm},
 and our algorithm for the integration over the time points and the summation of the diagrams is inspired by the method of modular path integrals \cite{makri2018modular}.
The following two sections will be devoted to these two steps, respectively.
\section{Inchworm Algorithm for Each Spin}
\label{sec_inchworm}
\label{Inchworm_with_cross}
Recall that our purpose is to compute the expectation of the observable in the form of  \cref{eq_observable}.
Based on our decomposition
\cref{full_propagator_diagrammatic},
 we can first take the trace for each diagram,
 and then sum up the results.
Thus, for each diagram, we need to compute $\tr\left( \rho_I^{(k)}(t) \mathcal{G}^{(k)}(\bs) \right)$ with $\rho_I^{(k)}(t) = \e^{-\ii H_0^{(k)} t} \rho^{(k)}(0) \e^{\ii H_0^{(k)} t}$.
In this section,
 we will introduce an efficient algorithm to evaluate this single-spin quantity $\mathcal{G}^{(k)}(\bs)$ for given $\bs$.
The algorithm is inspired by the inchworm Monte Carlo Method for system-bath coupling \cite{chen2017inchworm1,chen2017inchworm2},
where a single heat bath interacts with the entire system.

Note that each spin is associated with a thermal bath,
 we can apply the Dyson series expansion again to separate the spin and the bath.
Since the baths are initially in the thermal equilibrium states,
 the trace with respect to the bath part can be calculated explicitly using Wick's theorem \cite{wick1950evaluation}.
We refer the readers to \cite{chen2017inchworm1, cai2020inchworm} for the detailed calculation,
 and here we only present the final result:
\begin{equation}
\label{obsk}
    \tr \left(
    \rho_I^{(k)}(t) \mathcal{G}^{(k)}(\bs)
    \right)
    = \tr_{s^{(k)}} \hspace{-4pt}
    \left[
        \rho_{s,I}^{(k)}(t)
        \left(\prod_{n=1}^N \sqrt{\ii \, \sgn(s_n)} \right)
        \sum_{M=0}^{\infty} \ii^M \hspace{-4pt}
        \int_{-t \leqslant \boldsymbol{\tau} \leqslant t}
        \left( \prod_{m=1}^M \ii \, \sgn(\tau_m) \right)
            \mathcal{U}_0^{(k)}(\boldsymbol{\tau},\boldsymbol{s})
            \mathcal{L}_b^{(k)}(\boldsymbol{\tau})
        \dd \boldsymbol{\tau}
    \right],
\end{equation}
where
\begin{equation*}
    \mathcal{U}_0^{(k)}(\boldsymbol{\tau},\boldsymbol{s}) \\
    =  \mathcal{T}[V_{s,I}^{(k)}(s_1) \dots V_{s,I}^{(k)}(s_N) W_{s,I}^{(k)}(\tau_1) \dots W_{s,I}^{(k)}(\tau_M) O_{s,I}^{(k)}(0)]
\end{equation*}
with
\begin{equation*}
    V_{s,I}^{(k)}(s) 
    = \e^{-\ii H_s^{(k)} \vert s \vert} V^{(k)} 
    \e^{\ii H_s^{(k)} \vert s \vert}
    ,\quad
    W_{s,I}^{(k)}(\tau)
    =
    \e^{-\ii H_s^{(k)} \vert \tau \vert}
    W^{(k)}
    \e^{\ii H_s^{(k)} \vert \tau \vert},
    \quad
    \rho_{s,I}^{(k)}(t) = 
    \e^{-\ii H_s^{(k)} t} \rho_{s}^{(k)} (0) \e^{\ii H_s^{(k)} t} 
\end{equation*}
and the bath influence functional $\mathcal{L}_b^{(k)}(\boldsymbol{\tau})$ has the form \cite{Negele1988}
\begin{equation*}
    \mathcal{L}_b^{(k)}(\tau_1,\dots,\tau_M)
    = \begin{cases}
        0, &\text{if~} M \text{~is odd} \\
        \sum\limits_{\mathfrak{q}\in\mathcal{Q}_M}
        \prod\limits_{(j,j')\in \mathfrak{q}} B^{(k)}(\tau_j,\tau_{j'}),
        &\text{if~} M \text{~is even}.
    \end{cases}
\end{equation*}
Here $B^{(k)}$ is the two-point correlation function to be defined later in our test cases,
 and the set $\mathcal{Q}_M$ contains all possible pairings of integers $\{1,2,\cdots,M\}$.
For example,
\begin{equation*}
\begin{split}
    &\mathcal{Q}_2 = \{\{(1,2)\}\}, \\
    &\mathcal{Q}_4
    = \{ \{(1,2),(3,4)\}, \,
    \{(1,3),(2,4)\}, \,
    \{(1,4),(2,3)\}\}.
\end{split}
\end{equation*}
The general definition of $\mathcal{Q}_M$ for even $M$ is
\begin{equation}
    \label{Q}
    \mathcal{Q}_M = \left\{
    \left\{(j_1,j_1'),\dots,(j_{M/2},j_{M/2}')\right\}
    \Bigg\vert
    \bigcup_{l=1}^{M/2}\{j_l,j_l'\} = \{1,\dots,M\},
    j_l < j_l'\text{~for~} l = 1,\dots,M/2
    \right\},
\end{equation}
which includes $(M-1)!!$ pairings.

According to \cref{obsk}, now our objective is to evaluate the following quantity 
\begin{equation}
\label{eq_mathscrGkdefn}
    \mathscr{G}^{(k)}(-t,\boldsymbol{s},t)
    \coloneqq \left(\prod_{n=1}^N \sqrt{\ii \, \sgn(s_n)} \right)
    \sum_{M=0}^{\infty} 
    \int_{-t \leqslant \boldsymbol{\tau} \leqslant t}
    \left( \prod_{m=1}^M \ii \, \sgn(\tau_m) \right)
        \mathcal{U}_0^{(k)}(\boldsymbol{\tau},\boldsymbol{s})
        \mathcal{L}_b^{(k)}(\boldsymbol{\tau})
    \dd \boldsymbol{\tau},
\end{equation}
which yields  
\begin{equation}
\label{calG_scrG}
\tr\left(\rho_I^{(k)}(t) \mathcal{G}^{(k)}(\bs) \right) = \tr_{s^{(k)}}\left(\rho_{s,I}^{(k)}(t) \mathscr{G}^{(k)}(-t,\bs,t) \right).
\end{equation}
Recall that we have used a gray line with red crosses to represent $\mathcal{G}^{(k)}(\bs)$.
Due to the equivalence given in \cref{calG_scrG},
 below we will use the same diagram to represent the quantity $\mathscr{G}^{(k)}(-t,\bs,t)$.
For example, given $\bs = (s_1,s_2)$ with both $s_1$ and $s_2$ between $-t$ and $t$,
 \cref{eq_mathscrGkdefn} can be represented diagrammatically as
\begin{equation}
\label{bold_line}
\begin{split}
    \begin{tikzpicture}[baseline=0]
    \filldraw[fill=lightgray] (-2,-0.05) rectangle (2,0.05);
    \node[text=red] at (-1,0) {$\times$};
    \node[text=red] at (0.6,0) {$\times$};
    \node[text=black,anchor=north] at (-2,0) {$-t$};
    \node[text=black,anchor=north] at (2,0) {$t$};
    \node[text=black,anchor=north] at (-1,0) {$s_1$};
    \node[text=black,anchor=north] at (0.6,0) {$s_2$};
    \end{tikzpicture}
    &=
    \begin{tikzpicture}[baseline=0]
    \draw[black] (-2,0) -- (2,0);
    \node[text=red] at (-1,0) {$\times$};
    \node[text=red] at (0.6,0) {$\times$};
    \node[text=black,anchor=north] at (-2,0) {$-t$};
    \node[text=black,anchor=north] at (2,0) {$t$};
    \node[text=black,anchor=north] at (-1,0) {$s_1$};
    \node[text=black,anchor=north] at (0.6,0) {$s_2$};
    \end{tikzpicture}
    +
    \begin{tikzpicture}[baseline=0]
    \draw[black] (-2,0) -- (2,0);
    \node[text=red] at (-1,0) {$\times$};
    \node[text=red] at (0.6,0) {$\times$};
    \node[text=black,anchor=north] at (-2,0) {$-t$};
    \node[text=black,anchor=north] at (2,0) {$t$};
    \node[text=black,anchor=north] at (-1,0) {$s_1$};
    \node[text=black,anchor=north] at (0.6,0) {$s_2$};
    \draw[-] (-1.5,0) to[bend left=75] (-0.2,0);
    \node[text=black,anchor=north] at (-1.5,0) {$\tau_1$};
    \node[text=black,anchor=north] at (-0.2,0) {$\tau_2$};
    \end{tikzpicture} \\
    &+  
    \begin{tikzpicture}[baseline=0]
    \draw[black] (-2,0) -- (2,0);
    \node[text=red] at (-1,0) {$\times$};
    \node[text=red] at (0.6,0) {$\times$};
    \node[text=black,anchor=north] at (-2,0) {$-t$};
    \node[text=black,anchor=north] at (2,0) {$t$};
    \node[text=black,anchor=north] at (-1,0) {$s_1$};
    \node[text=black,anchor=north] at (0.6,0) {$s_2$};
    \draw[-] (-1.5,0) to[bend left=75] (-0.5,0);
    \draw[-] (0.1,0) to[bend left=75] (1.2,0);
    \node[text=black,anchor=north] at (-1.5,0) {$\tau_1$};
    \node[text=black,anchor=north] at (-0.5,0) {$\tau_2$};
    \node[text=black,anchor=north] at (0.1,0) {$\tau_3$};
    \node[text=black,anchor=north] at (1.2,0) {$\tau_4$};
    \end{tikzpicture}
    + 
    \begin{tikzpicture}[baseline=0]
    \draw[black] (-2,0) -- (2,0);
    \node[text=red] at (-1,0) {$\times$};
    \node[text=red] at (0.6,0) {$\times$};
    \node[text=black,anchor=north] at (-2,0) {$-t$};
    \node[text=black,anchor=north] at (2,0) {$t$};
    \node[text=black,anchor=north] at (-1,0) {$s_1$};
    \node[text=black,anchor=north] at (0.6,0) {$s_2$};
    \draw[-] (-1.5,0) to[bend left=75] (0.1,0);
    \draw[-] (-0.5,0) to[bend left=75] (1.2,0);
    \node[text=black,anchor=north] at (-1.5,0) {$\tau_1$};
    \node[text=black,anchor=north] at (-0.5,0) {$\tau_2$};
    \node[text=black,anchor=north] at (0.1,0) {$\tau_3$};
    \node[text=black,anchor=north] at (1.2,0) {$\tau_4$};
    \end{tikzpicture} + \dots
\end{split}
\end{equation}%
In the diagrammatic equation, the location of the cross marks, given by $\bs$, are fixed.
On the right hand side, $\tau$'s are integration variables. 
Note that a time ordering operator $\mathcal{T}$ in the definition of $\mathcal{U}_0^{(k)}(\boldsymbol{\tau}, \bs)$ is required to guarantee that the operators are applied in the correct order.
Each arc represents a two-point correlation function $B(\tau_j, \tau_{j'})$ in the bath influence functional $\mathcal{L}_b$.


The equation  \cref{bold_line} is ready for computation.
One can directly apply the Monte Carlo method to the right-hand side to approximate the sum of integrals,
 which is known as the bare diagrammatic quantum Monte Carlo method (bare dQMC).
To design a more efficient approach,
 we will follow the method in \cite{cai2020inchworm} to derive an integro-differential equation.
We first generalize the definition of $\mathscr{G}^{(k)}(-t,\boldsymbol{s},t)$ 
 to $\mathscr{G}^{(k)}(s_\ii, \bs, s_\ff)$ for any $s_\ii < s_\ff$: 
\begin{equation}
\label{eq_G_single_spin_generalized}
    \mathscr{G}^{(k)}(s_\ii,\boldsymbol{s},s_\ff)
    = \left( \prod_{n=1}^N \sqrt{\ii \, \sgn(s_n)} \right)
    \sum_{M=0}^{\infty} 
    \int_{s_\ii\leqslant \boldsymbol{\tau} \leqslant s_\ff}
    \left( \prod_{m=1}^M \ii \, \sgn(\tau_m) \right)
        \mathscr{U}_0^{(k)}(s_\ii,\boldsymbol{\tau},\boldsymbol{s},s_\ff) \mathcal{L}_b^{(k)}(\boldsymbol{\tau})
    \dd \boldsymbol{\tau}
\end{equation}
where $\bs$ is an increasing sequence of time points, each of which is between $s_\ii$ and $s_\ff$, and
\begin{equation*}
    \mathscr{U}_0^{(k)}(s_\ii,\boldsymbol{\tau},\boldsymbol{s},s_\ff)
    =\begin{cases}
        \mathcal{T}[V_{s,I}^{(k)}(s_1) \dots V_{s,I}^{(k)}(s_N) W_{s,I}^{(k)}(\tau_1) \dots W_{s,I}^{(k)}(\tau_M) O_s^{(k)}(0)],
        &\text{~if~} 0\in[s_\ii,s_\ff], \\
        \mathcal{T}[V_{s,I}^{(k)}(s_1) \dots V_{s,I}^{(k)}(s_N) W_{s,I}^{(k)}(\tau_1) \dots W_{s,I}^{(k)}(\tau_M)],
        &\text{~if~} 0\notin[s_\ii,s_\ff].
    \end{cases}
\end{equation*}
Note that only operators between $s_\ii$ and $s_\ff$ are included in the definition.
Therefore, when $[s_\ii, s_\ff]$ does not include the origin,
 $O_s(0)$ should be excluded.
This definition can also be represented diagramatically as  \cref{bold_line},
 only with $-t$ replaced by $s_\ii$ and $t$ replaced by $s_\ff$.
It can then be seen that for two intervals satisfying $[s_\ii, s_\ff] \subset [s_\ii', s_\ff']$,
 $\mathscr{G}^{(k)}(s_\ii, \bs, s_\ff)$ can be understood as a proportion of $\mathscr{G}^{(k)}(s_\ii', \bs', s_\ff')$ if $\bs$ is the subvector of $\bs'$ with all components between $s_\ii$ and $s_\ff$.

To formulate an integro-differential equation for $\mathscr{G}^{(k)}(s_\ii, \bs, s_\ff)$,
 we extend the gray line from $s_\ii$ to $s_\ff$ by a length of $\mathrm{d}s$ (see the left-hand side of \cref{bold_line_extended}).
Then in the expansion of the extended gray line,
 all diagrams on the right-hand side of \cref{bold_line} are included.
Besides, diagrams that are not included in \cref{bold_line} are thin lines with arcs ending within the interval $[s_\ff, s_\ff + \mathrm{d}s]$ (second line in \cref{bold_line_extended}).
Since $\mathrm{d}s$ is infinitesimal,
 it suffices to assume that there is only one time point inside $[s_\ff, s_\ff + \mathrm{d}s]$.
We can further assume that this time point is fixed at $s_\ff$,
 and then this diagram must be multiplied by $\mathrm{d}s$ when being added to the sum (third to fifth lines of \cref{bold_line_extended}). 
For simplicity,
 we will name the arc ending at $s_\ff$ as $\mathscr{A}_{s_\ff}$ (thick black arcs in \cref{bold_line_extended}).
We can now categorize all the diagrams with a point at $s_\ff$ into classes characterized by the connected component of the arcs including the arc $\mathscr{A}_{s_\ff}$.
Here the ``connected component'' can be established by beginning with a set including the arc $(\tau_k, \tau_M)$ only,
 and then expanding the set iteratively by including all arcs with intersections with any arc that is already in the set, until the set does not change.
In \cref{bold_line_extended},
 two categories are labeled by yellow and green backgrounds,
 and the connected components are highlighted using thick lines (including both black and white lines).
For all diagrams with the same connected component including $\mathscr{A}_{s_\ff}$,
 we can sum them up and the result is the connection of a few thick lines with all arcs in this connected component,
 which is known as a ``bold diagram''.
The derivation is summarized in the following diagrammatic equation:
\input{diagram_inchworm_derive.tex}%
where the notations $\tau$'s are omitted in some diagrams without ambiguity.
The mathematical formulae of the bold diagrams can be easily read off.
For example, the bold diagram with the yellow background should be interpreted as 
\begin{align*}
\highlightyellow{
    \int_{s_\ii}^{s_\ff} \dd \tau_1 
    \left(\ii \,\sgn(\tau_1)\, \ii\, \sgn(s_\ff)\right)
    W_s^{(k)}(s_\ff) 
    \mathscr{G}^{(k)}(\tau_1, \boldsymbol{s}_1 ,s_\ff)
    W_s^{(k)}(\tau_1)
    \mathscr{G}^{(k)}(s_\ii, \boldsymbol{s}_0 ,\tau_1)
    B^{(k)}(\tau_1,s_\ff),
}
\end{align*}
where $\boldsymbol{s}_0,\boldsymbol{s}_1$ are subsequences of $\boldsymbol{s}$ 
such that $(\boldsymbol{s}_0,\tau_1,\boldsymbol{s}_1)$ is an ascending sequence and  $\boldsymbol{s} = (\boldsymbol{s}_0,\boldsymbol{s}_1)$,
 and the bold diagram with the green background reads
\begin{align*}
    &\highlightgreen{\int_{s_\ii}^{s_\ff} \dd \tau_1 \dd \tau_2 \dd \tau_3 
    \left(\ii\, \sgn(\tau_1)\, \ii\, \sgn(\tau_2)\, \ii\, \sgn(\tau_3)\, \ii\, \sgn(s_\ff)\right)
    W_s^{(k)}(s_\ff) 
    \mathscr{G}^{(k)}(\tau_3, \boldsymbol{s}_3 ,s_\ff)
    } \\
    &\qquad\qquad \highlightgreen{W_s^{(k)}(\tau_1)
    \mathscr{G}^{(k)}(\tau_2, \boldsymbol{s}_2 ,\tau_3)
    W_s^{(k)}(\tau_1)
    \mathscr{G}^{(k)}(\tau_1, \boldsymbol{s}_1 ,\tau_2)
    W_s^{(k)}(\tau_1)
    \mathscr{G}^{(k)}(s_\ii, \boldsymbol{s}_0 ,\tau_1)
    B^{(k)}(\tau_1,\tau_3) B^{(k)}(\tau_2,s_\ff)
    }
\end{align*}
where $\boldsymbol{s}_0,\boldsymbol{s}_1,\boldsymbol{s}_2,\bs_3$ are subsequences of $\boldsymbol{s}$ such that
 $(\boldsymbol{s}_0,\tau_1,\boldsymbol{s}_1,\tau_2,\boldsymbol{s}_2,\tau_3,\boldsymbol{s}_3)$ is an ascending sequence and $\boldsymbol{s} = (\boldsymbol{s}_0,\boldsymbol{s}_1,\boldsymbol{s}_2,\boldsymbol{s}_3)$ .

The explicit expression of the diagrammatic equation \eqref{bold_line_extended} is as follows:
\begin{equation*}
  \mathscr{G}^{(k)}(s_\ii, \bs, s_\ff + \mathrm{d}s) = 
  \mathscr{G}^{(k)}(s_\ii, \bs, s_\ff)
  +
    \mathcal{K}^{(k)}(s_\ii, \bs, s_\ff) \,\mathrm{d}s,
\end{equation*}
where $\mathcal{K}^{(k)}(s_\ii, \bs, s_\ff)$ is the sum of bold diagrams inside the parentheses in \cref{bold_line_extended}:
\begin{displaymath}
    \mathcal{K}^{(k)}(s_\ii,\boldsymbol{s},s_\ff) 
    = 
        \begin{tikzpicture}[baseline=0,scale=0.8]
        \draw[black] (-2,0) -- (2,0);
        \filldraw[fill=lightgray] (-2,-0.05) rectangle (-0.012,0.05);
        \filldraw[fill=lightgray] (0.012,-0.05) rectangle (2,0.05);
        \node[text=red] at (-1.4,0) {$\times$};
        \node[text=red] at (-0.6,0) {$\times$};
        \node[text=red] at (1,0) {$\times$};
        \node[text=black,anchor=north] at (-2,0) {$s_\ii$};
        \node[text=black,anchor=north] at (2,0) {$s_\ff$};
        \draw[gray] (2,0.05) -- (2,-0.05);
        \node[text=black,anchor=north] at (-1.4,0) {$s_1$};
        \node[text=black,anchor=north] at (-0.6,0) {$s_2$};
        \node[text=black,anchor=north] at (0.2,-0.1) {$\dots$};
        \node[text=black,anchor=north] at (1,0) {$s_N$};
        \draw[-] (0,0) to[bend left=75] (2,0);
        \end{tikzpicture}
    + 
        \begin{tikzpicture}[baseline=0,scale=0.8]
        \draw[black] (-2,0) -- (2,0);
        \filldraw[fill=lightgray] (-2,-0.05) rectangle (-1.75-0.012,0.05);
        \filldraw[fill=lightgray] (-1.75+0.012,-0.05) rectangle (0-0.012,0.05);
        \filldraw[fill=lightgray] (0+0.012,-0.05) rectangle (0.25-0.012,0.05);
        \filldraw[fill=lightgray] (0.25+0.012,-0.05) rectangle (2,0.05);
        \node[text=red] at (-1.4,0) {$\times$};
        \node[text=red] at (-0.6,0) {$\times$};
        \node[text=red] at (1,0) {$\times$};
        \node[text=black,anchor=north] at (-2,0) {$s_\ii$};
        \node[text=black,anchor=north] at (2,0) {$s_\ff$};
        \draw[gray] (2,0.05) -- (2,-0.05);
        \node[text=black,anchor=north] at (-1.4,0) {$s_1$};
        \node[text=black,anchor=north] at (-0.6,0) {$s_2$};
        \node[text=black,anchor=north] at (0.2,-0.1) {$\dots$};
        \node[text=black,anchor=north] at (1,0) {$s_N$};
        \draw[-] (0,0) to[bend left=75] (2,0);
        \draw[-] (-1.75,0) to[bend left=75] (0.25,0);
        \end{tikzpicture}
    + \dots
\end{displaymath}
The integro-differential equation of $\mathscr{G}^{(k)}(s_\ii, \bs, s_\ff)$ can then be derived as
\begin{equation}
\label{inchworm_differential_equation}
    \pdv{\mathscr{G}^{(k)}(s_\ii,\boldsymbol{s},s_\ff)}{s_\ff}
    = \mathcal{K}^{(k)}(s_\ii,\bs,s_\ff).
\end{equation}
For the purpose of easier implementation,
 we will also provide the mathematical expression of $\mathcal{K}^{(k)}(s_\ii, \bs, s_\ff)$.
The general form of $\mathcal{K}^{(k)}(s_\ii, \bs, s_\ff)$ is
\begin{equation}
\label{eq_K}
    \mathcal{K}^{(k)}(s_\ii,\boldsymbol{s},s_\ff) 
    = \sum_{\substack{M=1\\M\text{~is~odd}}}^{\infty} 
    \int_{s_\ii \leqslant \tau_1 \leqslant \dots \leqslant \tau_{M} \leqslant s_\ff} 
    \dd \tau_1 \dots \dd \tau_M
    \left( \prod_{m=1}^{M+1} \ii \, \sgn (\tau_m) \right) W_s^{(k)}(s_\ff)
    \mathscr{U}^{(k)}(s_\ii,\boldsymbol{\tau},\boldsymbol{s},s_\ff) \mathcal{L}_b^{c(k)}(\boldsymbol{\tau}),
\end{equation}
where $\boldsymbol{\tau} = (\tau_1,\dots,\tau_M,\tau_{M+1})$ and $\tau_{M+1} = s_\ff$.
The system-associated operator $\mathscr{U}^{(k)}$ is defined by
\begin{equation}
\label{eq_U}
    \mathscr{U}^{(k)}(s_\ii,\boldsymbol{\tau},\boldsymbol{s},s_\ff)
    = \mathscr{G}^{(k)}(\tau_M, \bs_M, s_\ff) W_s^{(k)}(\tau_M) \mathscr{G}^{(k)}(\tau_{M-1}, \bs_{M-1}, \tau_M) W_s^{(k)}(\tau_{M-1}) \cdots W_s^{(k)}(\tau_1) \mathscr{G}^{(k)}(s_\ii, \bs_0, \tau_1)
\end{equation}
with $\bs_0, \cdots, \bs_M$ being subsequences of $\bs$ such that $\bs = (\bs_0, \bs_1, \cdots, \bs_M)$ 
and the extended sequence 
$$(s_\ii, \bs_0, \tau_1, \bs_1, \tau_2, \cdots, \tau_{M-1}, \bs_M, \tau_M)$$ 
is increasing. 
This indicates that $\bs_0, \cdots, \bs_M$ are subsequences of $\bs$ separated by $\tau_1, \cdots, \tau_M$.

The bath influence functional $\mathcal{L}_b^{c(k)}$ is exactly the same as the bath influence functional in \cite{cai2020inchworm}:
\begin{equation}
    \label{eq_L}
    \mathcal{L}_b^{c(k)}(\tau_1,\dots,\tau_{M+1})
    = \sum_{\mathfrak{q}\in\mathcal{Q}_{M+1}^c}
    \prod_{(j,{j'})\in\mathfrak{q}}
    B(\tau_j,\tau_{j'})
\end{equation}
where $\mathcal{Q}_{M+1}^c$ is the set of connected diagrams.
For example,
\begin{equation*}
\begin{split}
    &\mathcal{Q}_2^c = \{\{(1,2)\}\},  \\
    &\mathcal{Q}_4^c
    = \{\{(1,3),(2,4)\} \}, \\
    &\mathcal{Q}_6^c
    = \{
        \{(1,3),(2,5),(4,6)\},
        \{(1,4),(2,5),(3,6)\},
        \{(1,4),(2,6),(3,5)\},
        \{(1,5),(2,4),(3,6)\}
        \}.
\end{split}
\end{equation*}
One may refer to \cite[Section 3.3]{cai2020inchworm} for more information about the set $\mathcal{Q}_{M+1}^c$.
In general, the number of pairings in $\mathcal{Q}_{M+1}^c$ is asymptotically $\e^{-1} M!!$ when $M$ is a large odd integer \cite{stein1978class}.

For fixed $s_\ii$ and $\bs$, solving the integro-differential equation \cref{inchworm_differential_equation} requires an initial condition at $s_\ff = s_N$ (or $s_\ff = s_\ii$ if $\bs$ is an empty sequence). By definition, it can be immediately seen that
\begin{gather}
\label{init1}
\mathscr{G}^{(k)}(s_\ii, s_\ff = s_\ii) = \id^{(k)}, \qquad \text{if } s_\ii \neq 0, \\
\label{init2}
\mathscr{G}^{(k)}(s_\ii, s_1, \cdots, s_N, s_\ff = s_N) = \sqrt{\ii\, \sgn(s_N)} V_{s,I}^{(k)}(s_N) \mathscr{G}^{(k)}(s_\ii, s_1,\cdots,s_{N-1}, s_\ff = s_N), \qquad \text{if } s_N \neq 0.
\end{gather}
Due to the observable $O_s^{(k)}$ appearing in the definition of $\mathscr{G}^{(k)}$,
 there is a discontinuity when any of the time points touches zero.
The jump condition needed in the computation is
\begin{equation}
\label{jump}
  \lim_{s_\ff \rightarrow 0^+} \mathscr{G}^{(k)}(s_\ii,s_1,\dots,s_N,s_\ff) 
  = O_s^{(k)} 
  \lim_{s_\ff \rightarrow 0^-} \mathscr{G}^{(k)} (s_\ii,s_1,\dots,s_N,s_\ff).
\end{equation}
By these conditions,
 all the full propagators $\mathscr{G}^{(k)}(s_\ii, \bs, s_\ff)$ can be uniquely determined.

To solve the integro-differential equation  \eqref{inchworm_differential_equation} numerically,
 we start with solving all $\mathscr{G}^{(k)}(s_\ii, s_\ff)$, i.e. $N = 0$,
 and then increase the length of $\bs$ iteratively.
Such an order guarantees that the initial condition \cref{init2} can be applied whenever needed.
When solving $\mathscr{G}^{(k)}(s_\ii, \bs, s_\ff)$ for fixed $s_\ii$ and $\bs$,
 the second-order Heun's method is applied,
 and the jump condition \cref{jump} must be applied when $s_\ff$ crosses zero.
For the series of integrals on the right-hand side of \cref{inchworm_differential_equation},
 we select an odd positive integer $\bar{M}$ and truncate the series up to $M = \bar{M}$ as an approximation.
In our experiments,
 the value of $\bar{M}$ is at most $5$,
 and therefore the integrals in \cref{inchworm_differential_equation} are computed numerically using the second-order composite trapezoidal rule.
If larger $\bar{M}$ needs to be used,
 one can use Monte Carlo methods to approximate the integrals, leading to the inchworm Monte Carlo method as introduced in \cite{chen2017inchworm1, cai2020inchworm, cai2022numerical}.
To save computational cost, we have also utilized the following property of the full propagators: for all $T > 0$,
\begin{equation}
\label{eq_shift_invariance}
\begin{split}
    & \mathscr{G}^{(k)}(s_\ii+T,s_1+T,\dots,s_N+T,s_\ff+T) 
    = \e^{-\ii H_s T}
    \mathscr{G}^{(k)}(s_\ii,s_1,\dots,s_N,s_\ff)
    \e^{\ii H_s T},
    \text{~if~}s_\ii>0; \\
    & \mathscr{G}^{(k)}(s_\ii-T,s_1-T,\dots,s_N-T,s_\ff-T) 
    = \e^{-\ii H_s T}
    \mathscr{G}^{(k)}(s_\ii,s_1,\dots,s_N,s_\ff)
    \e^{\ii H_s T},
    \text{~if~}s_\ff<0.
\end{split}
\end{equation}
Note that the property holds only when all the time points are on the same side of the origin.
\section{Resummation of the full propagator}
\label{sec_resum}
Using the algorithm introduced in the previous section,
 we are able to compute all the gray lines in \cref{full_propagator_diagrammatic}.
In this section, 
 we will propose a fast algorithm to sum up all the diagrams.
Before introducing the algorithm,
 we first note that the same gray line for the same spin can sometimes be used multiple times during the summation.
For example,
 in the 4-spin case,
 when the propagator $\mathscr{G}^{(4)}(-t, s_1, t)$ is computed for the fourth spin,
 it can be applied in the following terms, all of which appear in \cref{full_propagator_diagrammatic}:
\begin{equation}
\label{last_spin_one_cross}
    \begin{tikzpicture}[baseline=0]
    \filldraw[fill=lightgray] (-1,0.55) rectangle (1,0.65);
    \filldraw[fill=lightgray] (-1,0.15) rectangle (1,0.25);
    \filldraw[fill=lightgray] (-1,-0.25) rectangle (1,-0.15);
    \filldraw[fill=lightgray] (-1,-0.65) rectangle (1,-0.55);
    \node[text=red] at (-0.5,-0.2) {$\times$};
    \node[text=red] at (-0.5,-0.6) {$\times$};
    \draw[black, densely dotted, line width = 1pt] (-0.5,-0.6) -- (-0.5,-0.2);
    \node[text=black,anchor=north] at (-1,-0.6) {$-t$};
    \node[text=black,anchor=north] at (1,-0.6) {$t$};
    \node[text=black,anchor=north] at (-0.5,-0.6) {$s_1$};
    \end{tikzpicture}, \quad
    \begin{tikzpicture}[baseline=0]
    \filldraw[fill=lightgray] (-1,0.55) rectangle (1,0.65);
    \filldraw[fill=lightgray] (-1,0.15) rectangle (1,0.25);
    \filldraw[fill=lightgray] (-1,-0.25) rectangle (1,-0.15);
    \filldraw[fill=lightgray] (-1,-0.65) rectangle (1,-0.55);
    \node[text=red] at (-0.5,0.6) {$\times$};
    \node[text=red] at (-0.5,0.2) {$\times$};
    \draw[black, densely dotted, line width = 1pt] (-0.5,0.6) -- (-0.5,0.2);
    \node[text=red] at (0.3,-0.2) {$\times$};
    \node[text=red] at (0.3,-0.6) {$\times$};
    \draw[black, densely dotted, line width = 1pt] (0.3,-0.2) -- (0.3,-0.6);
    \node[text=black,anchor=north] at (-1,-0.6) {$-t$};
    \node[text=black,anchor=north] at (1,-0.6) {$t$};
    \node[text=black,anchor=north] at (-0.5,-0.6) {$s_1$};
    \node[text=black,anchor=north] at (0.3,-0.6) {$s_2$};
    \end{tikzpicture}, \quad
    \begin{tikzpicture}[baseline=0]
    \filldraw[fill=lightgray] (-1,0.55) rectangle (1,0.65);
    \filldraw[fill=lightgray] (-1,0.15) rectangle (1,0.25);
    \filldraw[fill=lightgray] (-1,-0.25) rectangle (1,-0.15);
    \filldraw[fill=lightgray] (-1,-0.65) rectangle (1,-0.55);
    \node[text=red] at (-0.5,0.2) {$\times$};
    \node[text=red] at (-0.5,-0.2) {$\times$};
    \draw[black, densely dotted, line width = 1pt] (-0.5,0.2) -- (-0.5,-0.2);
    \node[text=red] at (0.3,-0.2) {$\times$};
    \node[text=red] at (0.3,-0.6) {$\times$};
    \draw[black, densely dotted, line width = 1pt] (0.3,-0.2) -- (0.3,-0.6);
    \node[text=black,anchor=north] at (-1,-0.6) {$-t$};
    \node[text=black,anchor=north] at (1,-0.6) {$t$};
    \node[text=black,anchor=north] at (-0.5,-0.6) {$s_1$};
    \node[text=black,anchor=north] at (0.3,-0.6) {$s_2$};
    \end{tikzpicture}, \quad
    \begin{tikzpicture}[baseline=0]
    \filldraw[fill=lightgray] (-1,0.55) rectangle (1,0.65);
    \filldraw[fill=lightgray] (-1,0.15) rectangle (1,0.25);
    \filldraw[fill=lightgray] (-1,-0.25) rectangle (1,-0.15);
    \filldraw[fill=lightgray] (-1,-0.65) rectangle (1,-0.55);
    \node[text=red] at (-0.5,-0.2) {$\times$};
    \node[text=red] at (-0.5,-0.6) {$\times$};
    \draw[black, densely dotted, line width = 1pt] (-0.5,-0.2) -- (-0.5,-0.6);
    \node[text=red] at (0.3,0.6) {$\times$};
    \node[text=red] at (0.3,0.2) {$\times$};
    \draw[black, densely dotted, line width = 1pt] (0.3,0.6) -- (0.3,0.2);
    \node[text=black,anchor=north] at (-1,-0.6) {$-t$};
    \node[text=black,anchor=north] at (1,-0.6) {$t$};
    \node[text=black,anchor=north] at (-0.5,-0.6) {$s_1$};
    \node[text=black,anchor=north] at (0.3,-0.6) {$s_2$};
    \end{tikzpicture}, \quad
    \begin{tikzpicture}[baseline=0]
    \filldraw[fill=lightgray] (-1,0.55) rectangle (1,0.65);
    \filldraw[fill=lightgray] (-1,0.15) rectangle (1,0.25);
    \filldraw[fill=lightgray] (-1,-0.25) rectangle (1,-0.15);
    \filldraw[fill=lightgray] (-1,-0.65) rectangle (1,-0.55);
    \node[text=red] at (-0.5,-0.2) {$\times$};
    \node[text=red] at (-0.5,-0.6) {$\times$};
    \draw[black, densely dotted, line width = 1pt] (-0.5,-0.2) -- (-0.5,-0.6);
    \node[text=red] at (0.3,0.2) {$\times$};
    \node[text=red] at (0.3,-0.2) {$\times$};
    \draw[black, densely dotted, line width = 1pt] (0.3,0.2) -- (0.3,-0.2);
    \node[text=black,anchor=north] at (-1,-0.6) {$-t$};
    \node[text=black,anchor=north] at (1,-0.6) {$t$};
    \node[text=black,anchor=north] at (-0.5,-0.6) {$s_1$};
    \node[text=black,anchor=north] at (0.3,-0.6) {$s_2$};
    \end{tikzpicture}.
\end{equation}
Instead of applying \cref{full_propagator_diagrammatic} directly to compute the summation, we will follow the idea of the modular path integral \cite{makri2018modular} to assemble all the gray lines by adding spins iteratively.

Assuming that we want to add up all the five diagrams in \cref{last_spin_one_cross}.
Notice that the terms related to the last spin are essentially the same in all these diagrams.
Therefore, instead of computing all the diagrams,
 a more efficient way is to apply the distributive law to separate the last spin and only add up the terms for the first three spins.
Similarly, when dealing with the sum involving the first three spins,
 the first and the second diagrams in \cref{last_spin_one_cross} can be combined;
 the third and the fifth diagrams in \cref{last_spin_one_cross} can also be combined.
In general,
 to deal with the sum on the right-hand side of  \cref{full_propagator_diagrammatic},
 we can first separate all the diagrams in to groups according to the number of crosses on the last line.
Then, for each of the groups,
 we further separate the diagrams into subgroups according to the crosses on third line.
For each of the subgroups,
 we apply such grouping one more time according to the crosses on the second line.
When performing computations,
 we first sum up the terms involving only the first spin in all the smallest groups.
For the result of each group,
 we multiply them by the corresponding term related to the second spin,
 and then repeat a similar procedure for rest of the spins.
Mathmatically, this idea is based on the following iterative representation of the observable:
\begin{align}
    \label{first_spin}
    &G^{[1]}(-t,\boldsymbol{s},t) = 
    \tr_{s^{(1)}} \left(
    \rho_{s,I}^{(1)}(t)
    \mathscr{G}^{(1)}(-t,\boldsymbol{s},t) \right); \\
    \label{add_spin}
    &G^{[k+1]}(-t,\boldsymbol{s},t)
    = \sum_{N' = 0}^{\infty} 
    \int_{-t\leqslant \boldsymbol{s'} \leqslant t}
    G^{[k]}\left(-t,\boldsymbol{s'},t\right)
    \tr_{s^{(k+1)}}\left(
    \rho_{s,I}^{(k+1)}(t)
    \mathscr{G}^{(k+1)}(-t,\mathcal{P}(\boldsymbol{s},\boldsymbol{s'}),t)
    \dd \boldsymbol{s'}\right), \\
    &\hspace{300pt}
    \text{~for~} k = 1,\dots,n-2; \notag \\
    \label{close_diagram}
    &G^{[K]}(-t,t) = \sum_{N=0}^{\infty}
    \int_{-t\leqslant \boldsymbol{s} \leqslant t}
    G^{[K-1]}(-t,\boldsymbol{s},t) 
    \tr_{s^{(K)}}
    \left(
    \rho_{s,I}^{(K)}(t)
    \mathscr{G}^{(K)}(-t,\boldsymbol{s},t) \right)
    \dd \boldsymbol{s}
\end{align}
where $\boldsymbol{s'}=(s'_1,\dots,s'_{N'})$ and 
$\boldsymbol{s}=(s_1,\dots,s_{N})$ are two non-descending lists.
In \cref{add_spin}, $\mathcal{P}$ is the sorting operator to merge $\boldsymbol{s}$ and $\boldsymbol{s'}$ into a sorted list.
We start from the first spin with \cref{first_spin}, add the middle spins by \cref{add_spin} and close the diagram by \cref{close_diagram}.
These equations show that there are many duplicate computations in the procedure above,
 which can be avoided.
The details of the final algorithm will again be illustrated using diagrams below.

The computation of \eqref{first_spin} is straightforward.
We start our discussion with the case $j = 1$ in \cref{add_spin},
 which becomes
\begin{equation}
    G^{[2]}(-t,\boldsymbol{s},t)
    = \sum_{N' = 0}^{\infty} 
    \int_{-t\leqslant \boldsymbol{s'} \leqslant t}
    G^{[1]}\left(-t,\mathcal{P}(\boldsymbol{s},\boldsymbol{s'}),t\right)
    \tr_{s^{(2)}} \left( \rho_{s,I}^{(2)}(t) \mathscr{G}^{(2)}(-t,\boldsymbol{s}',t) \right)
    \dd \boldsymbol{s}'.
\end{equation}
If $\boldsymbol{s}$ has length 1,
 the equation can be diagrammatically represented by
\begin{equation}
\label{G2}
    G^{[2]}(-t,s_1,t) = 
    \begin{tikzpicture}[baseline=0]
        \filldraw[fill=lightgray] (-1,0.55-0.4) rectangle (1,0.65-0.4);
        \filldraw[fill=lightgray] (-1,0.15-0.4) rectangle (1,0.25-0.4);
        \draw[black, line width=1.5pt] (-1,0.65-0.4) -- (-1,0.15-0.4);
        \draw[black, line width=1.5pt] (+1,0.65-0.4) -- (+1,0.15-0.4);
        \node[text=red] at (-0.5,0.2-0.4) {$\times$};
        \draw[black, densely dotted, line width = 1pt] (-0.5,0.2-0.4) -- (-0.5,-0.2-0.4);
    \end{tikzpicture}
    = 
    \begin{tikzpicture}[baseline=0]
        \filldraw[fill=lightgray] (-1,0.55-0.4) rectangle (1,0.65-0.4);
        \filldraw[fill=lightgray] (-1,0.15-0.4) rectangle (1,0.25-0.4);
        \node[text=red] at (-0.5,0.2-0.4) {$\times$};
        \draw[black, densely dotted, line width = 1pt] (-0.5,0.2-0.4) -- (-0.5,-0.2-0.4);
    \end{tikzpicture}
    +
    \begin{tikzpicture}[baseline=0]
        \filldraw[fill=lightgray] (-1,0.55-0.4) rectangle (1,0.65-0.4);
        \filldraw[fill=lightgray] (-1,0.15-0.4) rectangle (1,0.25-0.4);
        \node[text=red] at (-0.5,0.2-0.4) {$\times$};
        \draw[black, densely dotted, line width = 1pt] (-0.5,0.2-0.4) -- (-0.5,-0.2-0.4);
        \node[text=red] at (0.2,0.6-0.4) {$\times$};
        \node[text=red] at (0.2,0.2-0.4) {$\times$};
        \draw[black, densely dotted, line width = 1pt] (0.2,0.6-0.4) -- (0.2,0.2-0.4);
    \end{tikzpicture}
    +
    \begin{tikzpicture}[baseline=0]
        \filldraw[fill=lightgray] (-1,0.55-0.4) rectangle (1,0.65-0.4);
        \filldraw[fill=lightgray] (-1,0.15-0.4) rectangle (1,0.25-0.4);
        \node[text=red] at (-0.5,0.2-0.4) {$\times$};
        \draw[black, densely dotted, line width = 1pt] (-0.5,0.2-0.4) -- (-0.5,-0.2-0.4);
        \node[text=red] at (0.1,0.6-0.4) {$\times$};
        \node[text=red] at (0.1,0.2-0.4) {$\times$};
        \draw[black, densely dotted, line width = 1pt] (0.1,0.6-0.4) -- (0.1,0.2-0.4);
        \node[text=red] at (-0.8,0.6-0.4) {$\times$};
        \node[text=red] at (-0.8,0.2-0.4) {$\times$};
        \draw[black, densely dotted, line width = 1pt] (-0.8,0.6-0.4) -- (-0.8,0.2-0.4);
    \end{tikzpicture}
    + 
    \begin{tikzpicture}[baseline=0]
        \filldraw[fill=lightgray] (-1,0.55-0.4) rectangle (1,0.65-0.4);
        \filldraw[fill=lightgray] (-1,0.15-0.4) rectangle (1,0.25-0.4);
        \node[text=red] at (-0.5,0.2-0.4) {$\times$};
        \draw[black, densely dotted, line width = 1pt] (-0.5,0.2-0.4) -- (-0.5,-0.2-0.4);
        \node[text=red] at (0.1,0.6-0.4) {$\times$};
        \node[text=red] at (0.1,0.2-0.4) {$\times$};
        \draw[black, densely dotted, line width = 1pt] (0.1,0.6-0.4) -- (0.1,0.2-0.4);
        \node[text=red] at (-0.8,0.6-0.4) {$\times$};
        \node[text=red] at (-0.8,0.2-0.4) {$\times$};
        \draw[black, densely dotted, line width = 1pt] (-0.8,0.6-0.4) -- (-0.8,0.2-0.4);
        \node[text=red] at (0.7,0.6-0.4) {$\times$};
        \node[text=red] at (0.7,0.2-0.4) {$\times$};
        \draw[black, densely dotted, line width = 1pt] (0.7,0.6-0.4) -- (0.7,0.2-0.4);
    \end{tikzpicture}
    + \dots.
\end{equation}%
On the left-hand side,
 the diagram represents the quantity $G^{[2]}(-t,\boldsymbol{s},t)$ where the two short black lines binding the bold lines indicate that all connections between the first two spins are taken into account.
The parameter $\boldsymbol{s}$ is shown as the cross on the second spin.
We use an open dashed line to indicate that it will be connected to the third spin in the next step.
The right-hand side of the equation represents the sum and the integral in \cref{G2}.
The four diagrams represent the terms for $N' = 0,1,2,3,4$, respectively.
Similarly, if the length of $\boldsymbol{s}$ is 2, we have the following diagrammatic equation:
\begin{equation*}
    G^{[2]}(-t,s_1,s_2,t) =
    \begin{tikzpicture}[baseline=0]
        \filldraw[fill=lightgray] (-1,0.55-0.4) rectangle (1,0.65-0.4);
        \filldraw[fill=lightgray] (-1,0.15-0.4) rectangle (1,0.25-0.4);
        \draw[black, line width=1.5pt] (-1,0.65-0.4) -- (-1,0.15-0.4);
        \draw[black, line width=1.5pt] (+1,0.65-0.4) -- (+1,0.15-0.4);
        \node[text=red] at (-0.5,0.2-0.4) {$\times$};
        \draw[black, densely dotted, line width=1pt] (-0.5,0.2-0.4) -- (-0.5,-0.2-0.4);
        \node[text=red] at (0.5,0.2-0.4) {$\times$};
        \draw[black, densely dotted, line width=1pt] (0.5,0.2-0.4) -- (0.5,-0.2-0.4);
    \end{tikzpicture}
    = 
    \begin{tikzpicture}[baseline=0]
        \filldraw[fill=lightgray] (-1,0.55-0.4) rectangle (1,0.65-0.4);
        \filldraw[fill=lightgray] (-1,0.15-0.4) rectangle (1,0.25-0.4);
        \node[text=red] at (-0.5,0.2-0.4) {$\times$};
        \draw[black, densely dotted, line width=1pt] (-0.5,0.2-0.4) -- (-0.5,-0.2-0.4);
        \node[text=red] at (0.5,0.2-0.4) {$\times$};
        \draw[black, densely dotted, line width=1pt] (0.5,0.2-0.4) -- (0.5,-0.2-0.4);
    \end{tikzpicture}
    +
    \begin{tikzpicture}[baseline=0]
        \filldraw[fill=lightgray] (-1,0.55-0.4) rectangle (1,0.65-0.4);
        \filldraw[fill=lightgray] (-1,0.15-0.4) rectangle (1,0.25-0.4);
        \node[text=red] at (-0.5,0.2-0.4) {$\times$};
        \draw[black, densely dotted, line width=1pt] (-0.5,0.2-0.4) -- (-0.5,-0.2-0.4);
        \node[text=red] at (0.2,0.6-0.4) {$\times$};
        \node[text=red] at (0.2,0.2-0.4) {$\times$};
        \draw[black, densely dotted, line width=1pt] (0.2,0.6-0.4) -- (0.2,0.2-0.4);
        \node[text=red] at (0.5,0.2-0.4) {$\times$};
        \draw[black, densely dotted, line width=1pt] (0.5,0.2-0.4) -- (0.5,-0.2-0.4);
    \end{tikzpicture}
    +
    \begin{tikzpicture}[baseline=0]
        \filldraw[fill=lightgray] (-1,0.55-0.4) rectangle (1,0.65-0.4);
        \filldraw[fill=lightgray] (-1,0.15-0.4) rectangle (1,0.25-0.4);
        \node[text=red] at (-0.5,0.2-0.4) {$\times$};
        \draw[black, densely dotted, line width=1pt] (-0.5,0.2-0.4) -- (-0.5,-0.2-0.4);
        \node[text=red] at (0.1,0.6-0.4) {$\times$};
        \node[text=red] at (0.1,0.2-0.4) {$\times$};
        \draw[black, densely dotted, line width=1pt] (0.1,0.6-0.4) -- (0.1,0.2-0.4);
        \node[text=red] at (-0.8,0.6-0.4) {$\times$};
        \node[text=red] at (-0.8,0.2-0.4) {$\times$};
        \draw[black, densely dotted, line width=1pt] (-0.8,0.6-0.4) -- (-0.8,0.2-0.4);
        \node[text=red] at (0.5,0.2-0.4) {$\times$};
        \draw[black, densely dotted, line width=1pt] (0.5,0.2-0.4) -- (0.5,-0.2-0.4);
    \end{tikzpicture}
    + 
    \begin{tikzpicture}[baseline=0]
        \filldraw[fill=lightgray] (-1,0.55-0.4) rectangle (1,0.65-0.4);
        \filldraw[fill=lightgray] (-1,0.15-0.4) rectangle (1,0.25-0.4);
        \node[text=red] at (-0.5,0.2-0.4) {$\times$};
        \draw[black, densely dotted, line width=1pt] (-0.5,0.2-0.4) -- (-0.5,-0.2-0.4);
        \node[text=red] at (0.1,0.6-0.4) {$\times$};
        \node[text=red] at (0.1,0.2-0.4) {$\times$};
        \draw[black, densely dotted, line width=1pt] (0.1,0.6-0.4) -- (0.1,0.2-0.4);
        \node[text=red] at (-0.8,0.6-0.4) {$\times$};
        \node[text=red] at (-0.8,0.2-0.4) {$\times$};
        \draw[black, densely dotted, line width=1pt] (-0.8,0.6-0.4) -- (-0.8,0.2-0.4);
        \node[text=red] at (0.7,0.6-0.4) {$\times$};
        \node[text=red] at (0.7,0.2-0.4) {$\times$};
        \draw[black, densely dotted, line width=1pt] (0.7,0.6-0.4) -- (0.7,0.2-0.4);
        \node[text=red] at (0.5,0.2-0.4) {$\times$};
        \draw[black, densely dotted, line width=1pt] (0.5,0.2-0.4) -- (0.5,-0.2-0.4);
    \end{tikzpicture}
    + \dots.
\end{equation*}

After computing the values of $G^{[2]}(-t,\boldsymbol{s},t)$
 for all $\boldsymbol{s}$,
 we can move forward to adding the third spin into the diagram.
An example for $N=3$ is
\begin{displaymath}
    G^{[3]}(-t,s_1,s_2,s_3,t) =
    \begin{tikzpicture}[baseline=0]
        \filldraw[fill=lightgray] (-1,0.55) rectangle (1,0.65);
        \filldraw[fill=lightgray] (-1,0.15) rectangle (1,0.25);
        \filldraw[fill=lightgray] (-1,-0.15) rectangle (1,-0.25);
        \draw[black, line width=1.5pt] (-1,0.65) -- (-1,-0.25);
        \draw[black, line width=1.5pt] (+1,0.65) -- (+1,-0.25);
        \node[text=red] at (-0.5,0.2-0.4) {$\times$};
        \draw[black, densely dotted, line width=1pt] (-0.5,0.2-0.4) -- (-0.5,-0.2-0.4);
        \node[text=red] at (0.2,0.2-0.4) {$\times$};
        \draw[black, densely dotted, line width=1pt] (0.2,0.2-0.4) -- (0.2,-0.2-0.4);
        \node[text=red] at (0.7,0.2-0.4) {$\times$};
        \draw[black, densely dotted, line width=1pt] (0.7,0.2-0.4) -- (0.7,-0.2-0.4);
    \end{tikzpicture} =
    \begin{tikzpicture}[baseline=0]
        \filldraw[fill=lightgray] (-1,0.55) rectangle (1,0.65);
        \filldraw[fill=lightgray] (-1,0.15) rectangle (1,0.25);
        \filldraw[fill=lightgray] (-1,-0.15) rectangle (1,-0.25);
        \draw[black, line width=1.5pt] (-1,0.65) -- (-1,-0.25+0.4);
        \draw[black, line width=1.5pt] (+1,0.65) -- (+1,-0.25+0.4);
        \node[text=red] at (-0.5,0.2-0.4) {$\times$};
        \draw[black, densely dotted, line width=1pt] (-0.5,0.2-0.4) -- (-0.5,-0.2-0.4);
        \node[text=red] at (0.2,0.2-0.4) {$\times$};
        \draw[black, densely dotted, line width=1pt] (0.2,0.2-0.4) -- (0.2,-0.2-0.4);
        \node[text=red] at (0.7,0.2-0.4) {$\times$};
        \draw[black, densely dotted, line width=1pt] (0.7,0.2-0.4) -- (0.7,-0.2-0.4);
    \end{tikzpicture} +
    \begin{tikzpicture}[baseline=0]
        \filldraw[fill=lightgray] (-1,0.55) rectangle (1,0.65);
        \filldraw[fill=lightgray] (-1,0.15) rectangle (1,0.25);
        \filldraw[fill=lightgray] (-1,-0.15) rectangle (1,-0.25);
        \draw[black, line width=1.5pt] (-1,0.65) -- (-1,-0.25+0.4);
        \draw[black, line width=1.5pt] (+1,0.65) -- (+1,-0.25+0.4);
        \node[text=red] at (-0.5,0.2-0.4) {$\times$};
        \draw[black, densely dotted, line width=1pt] (-0.5,0.2-0.4) -- (-0.5,-0.2-0.4);
        \node[text=red] at (0.2,0.2-0.4) {$\times$};
        \draw[black, densely dotted, line width=1pt] (0.2,0.2-0.4) -- (0.2,-0.2-0.4);
        \node[text=red] at (0.7,0.2-0.4) {$\times$};
        \draw[black, densely dotted, line width=1pt] (0.7,0.2-0.4) -- (0.7,-0.2-0.4);
        \node[text=red] at (0.4,0.2) {$\times$};
        \draw[black, densely dotted, line width=1pt] (0.4,0.2) -- (0.4,-0.2);
        \node[text=red] at (0.4,0.2-0.4) {$\times$};
    \end{tikzpicture} +
    \begin{tikzpicture}[baseline=0]
        \filldraw[fill=lightgray] (-1,0.55) rectangle (1,0.65);
        \filldraw[fill=lightgray] (-1,0.15) rectangle (1,0.25);
        \filldraw[fill=lightgray] (-1,-0.15) rectangle (1,-0.25);
        \draw[black, line width=1.5pt] (-1,0.65) -- (-1,-0.25+0.4);
        \draw[black, line width=1.5pt] (+1,0.65) -- (+1,-0.25+0.4);
        \node[text=red] at (-0.5,0.2-0.4) {$\times$};
        \draw[black, densely dotted, line width=1pt] (-0.5,0.2-0.4) -- (-0.5,-0.2-0.4);
        \node[text=red] at (0.2,0.2-0.4) {$\times$};
        \draw[black, densely dotted, line width=1pt] (0.2,0.2-0.4) -- (0.2,-0.2-0.4);
        \node[text=red] at (0.7,0.2-0.4) {$\times$};
        \draw[black, densely dotted, line width=1pt] (0.7,0.2-0.4) -- (0.7,-0.2-0.4);
        \node[text=red] at (0.4,0.2) {$\times$};
        \draw[black, densely dotted, line width=1pt] (0.4,0.2) -- (0.4,-0.2);
        \node[text=red] at (0.4,0.2-0.4) {$\times$};
        \node[text=red] at (-0.2,0.2) {$\times$};
        \draw[black, densely dotted, line width=1pt] (-0.2,0.2) -- (-0.2,-0.2);
        \node[text=red] at (-0.2,0.2-0.4) {$\times$};
    \end{tikzpicture} + \cdots
\end{displaymath}
We then repeat this process recurrently until we add the second last spin into the diagram. This completes the computation of \cref{add_spin}.

To add the last spin, \cref{close_diagram} is applied instead of \cref{add_spin}.
The only difference is that there are no further spins so that the time sequence $\boldsymbol{s}$ in $G^{[K]}(-t,\boldsymbol{s},t)$ can only be an empty list,
 which will then be simply denoted by $G^{[K]}(-t,t)$.
Diagrammatically, in the 4-spin case, the last step can be represented by
\begin{equation*}
    \begin{tikzpicture}[baseline=0]
        \filldraw[fill=lightgray] (-1,0.55) rectangle (1,0.65);
        \filldraw[fill=lightgray] (-1,0.15) rectangle (1,0.25);
        \filldraw[fill=lightgray] (-1,-0.55) rectangle (1,-0.65);
        \filldraw[fill=lightgray] (-1,-0.15) rectangle (1,-0.25);
        \draw[black, line width=1.5pt] (-1,0.65) -- (-1,-0.65);
        \draw[black, line width=1.5pt] (+1,0.65) -- (+1,-0.65);
    \end{tikzpicture}
    = 
    \begin{tikzpicture}[baseline=0]
        \filldraw[fill=lightgray] (-1,0.55) rectangle (1,0.65);
        \filldraw[fill=lightgray] (-1,0.15) rectangle (1,0.25);
        \filldraw[fill=lightgray] (-1,-0.15) rectangle (1,-0.25);
        \filldraw[fill=lightgray] (-1,-0.55) rectangle (1,-0.65);
        \draw[black, line width=1.5pt] (-1,0.65) -- (-1,-0.25);
        \draw[black, line width=1.5pt] (+1,0.65) -- (+1,-0.25);
    \end{tikzpicture}
    +
    \begin{tikzpicture}[baseline=0]
        \filldraw[fill=lightgray] (-1,0.55) rectangle (1,0.65);
        \filldraw[fill=lightgray] (-1,0.15) rectangle (1,0.25);
        \filldraw[fill=lightgray] (-1,-0.15) rectangle (1,-0.25);
        \filldraw[fill=lightgray] (-1,-0.55) rectangle (1,-0.65);
        \draw[black, line width=1.5pt] (-1,0.65) -- (-1,-0.25);
        \draw[black, line width=1.5pt] (+1,0.65) -- (+1,-0.25);
        \node[text=red] at (-0.5,0.2-0.4) {$\times$};
        \node[text=red] at (-0.5,-0.2-0.4) {$\times$};
        \draw[black, densely dotted, line width=1pt] (-0.5,0.2-0.4) -- (-0.5,-0.2-0.4);
    \end{tikzpicture}
    +
    \begin{tikzpicture}[baseline=0]
        \filldraw[fill=lightgray] (-1,0.55) rectangle (1,0.65);
        \filldraw[fill=lightgray] (-1,0.15) rectangle (1,0.25);
        \filldraw[fill=lightgray] (-1,-0.15) rectangle (1,-0.25);
        \filldraw[fill=lightgray] (-1,-0.55) rectangle (1,-0.65);
        \draw[black, line width=1.5pt] (-1,0.65) -- (-1,-0.25);
        \draw[black, line width=1.5pt] (+1,0.65) -- (+1,-0.25);
        \node[text=red] at (-0.5,0.2-0.4) {$\times$};
        \node[text=red] at (-0.5,-0.2-0.4) {$\times$};
        \draw[black, densely dotted, line width=1pt] (-0.5,0.2-0.4) -- (-0.5,-0.2-0.4);
        \node[text=red] at (0.2,0.2-0.4) {$\times$};
        \node[text=red] at (0.2,-0.2-0.4) {$\times$};
        \draw[black, densely dotted, line width=1pt] (0.2,0.2-0.4) -- (0.2,-0.2-0.4);
    \end{tikzpicture}
    +
    \begin{tikzpicture}[baseline=0]
        \filldraw[fill=lightgray] (-1,0.55) rectangle (1,0.65);
        \filldraw[fill=lightgray] (-1,0.15) rectangle (1,0.25);
        \filldraw[fill=lightgray] (-1,-0.15) rectangle (1,-0.25);
        \filldraw[fill=lightgray] (-1,-0.55) rectangle (1,-0.65);
        \draw[black, line width=1.5pt] (-1,0.65) -- (-1,-0.25);
        \draw[black, line width=1.5pt] (+1,0.65) -- (+1,-0.25);
        \node[text=red] at (-0.5,0.2-0.4) {$\times$};
        \node[text=red] at (-0.5,-0.2-0.4) {$\times$};
        \draw[black, densely dotted, line width=1pt] (-0.5,0.2-0.4) -- (-0.5,-0.2-0.4);
        \node[text=red] at (0.2,0.2-0.4) {$\times$};
        \node[text=red] at (0.2,-0.2-0.4) {$\times$};
        \draw[black, densely dotted, line width=1pt] (0.2,0.2-0.4) -- (0.2,-0.2-0.4);
        \node[text=red] at (0.7,0.2-0.4) {$\times$};
        \node[text=red] at (0.7,-0.2-0.4) {$\times$};
        \draw[black, densely dotted, line width=1pt] (0.7,0.2-0.4) -- (0.7,-0.2-0.4);
    \end{tikzpicture}
    + \dots
\end{equation*}%
Additionally, the quantity of the left hand side is exactly the quantity $\expval{O_s} = \tr \left(\rho_I(t) G(-t,t)\right)$.

In practical simulations, it is impossible to consider an infinite number of diagrams.
Instead, a sufficiently large integer $\bar{N}$ is chosen as the maximum number of interactions between any spin and its neighboring spins. 
Diagrammatically, $\bar{N}$ corresponds to the maximum number of red crosses on each line. Furthermore, as depicted in \cref{add_spin}, each diagram corresponds to an integral over a simplex, 
 which is approximated using the composite trapezoidal quadrature rule in our numerical implementation.
Recall that the integro-differential equation is also solved using a second-order method.
The overall convergence rate of our method is second order.

\begin{remark}
\label{remark_MPI}
Here we would like to comment the relation and difference between modular path integral (MPI) proposed in \cite{makri2018communication,makri2018modular} and our approach.
Both methods compute the Ising chain dynamics iteratively based on the connection of spins.
MPI utilizes QuAPI for the computation of a single spin dynamics 
while our method uses the Inchworm algorithm.
Another significant difference between two methods is that MPI considers all possible connections between spins 
given a specific time discretization
while our method, instead, introduces a cut off for the spin couplings.
With the cut-off,
 it is possible to reduce the number of diagrams
 and hence improve the computational efficiency.
\end{remark}
\section{Estimation of the Computational Cost}
\label{sec_computational_cost}
In this section,
 we estimate the computational cost for our method.
As discussed above, the computation contains two parts,
 including the computation of all bold lines with red crosses for all the spins (\cref{sec_inchworm})
 and the summation of the full propagators (\cref{sec_resum}).
For simplicity,
 a uniform time step $\dt$ is chosen throughout the computation.
All the discrete time points are therefore multiples of $\dt$.
Below we will estimate the cost for computing $G(-t,t)$ for $t=\dt,2\dt,\dots,L\dt$ given a positive integer $L$.

\subsection{Computational cost for each spin}
\label{sec_computational_cost_one_spin}
The integro-differential equation \eqref{inchworm_differential_equation} shows that the computation of longer diagrams depends on the knowledge of shorter diagrams.
To compute $G(-t,t)$ for $t$ up to $L\dt$,
 the maximum length of the diagrams is $2L\dt$.
For any $l = 1,\cdots,2L$, we can then assume that all the diagrams of length less than $l \Delta t$ are already computed,
 and focus on the diagrams of length $l \Delta t$.
For fixed $l$, the computational costs for all diagrams of length $l\Delta t$ are generally the same.
The most costly part is the computation of $\mathcal{G}^{(k)}(s_\ii,\boldsymbol{s},s_\ff)$ in \cref{inchworm_differential_equation}.
Taking the forward Euler method as an example,
 we need to evaluate $\mathcal{K}^{(k)}(s_\ii,\boldsymbol{s},s_\ii + (l-1)\dt)$ to obtain $\mathcal{G}^{(k)}(s_\ii,\boldsymbol{s},s_\ii + l \dt)$.
According to  \cref{eq_K},
 the computational cost can be estimated by
\begin{equation}
\label{cost_for_K}
    \sum_{\substack{M=1\\M \text{ is odd}}}^{\bar{M}} C_M \genfrac(){0pt}{0}{M+l}{M},
\end{equation}
where the binomial coefficient $\genfrac(){0pt}{2}{M+l}{M}$ is the number of grid points in the $M$-dimensional simplex $s_\ii \leqslant \bs \leqslant s_\ii + (l-1)\dt$, and $C_M$ is the computational cost of the integrand.
Note that this estimation is based on the grid-based numerical quadrature,
 which does not apply to Monte Carlo methods.
For large $M$,
 the computation of the bath influence functional becomes dominant since the number of diagrams increases as $\mathcal{O}(M!!)$,
 so that $C_M$ can be estimated by $\mathcal{O}((M+2)!!)$.
In our tests,
 $\bar{M}$ is no more than $5$.
Hence, we will regard $C_M$ as a constant for simplicity.

With the cost of each diagram estimated by \cref{cost_for_K},
 we now need to calculate the number of diagrams of length $l\dt$.
The estimation of the computational cost starts from the number of different bold-lines with total length $l\dt$
for $l=1,\dots,2L$.
When $l \leqslant L$,
 the interval $[s_\ii,s_\ff]$ may or may not contain the origin 0.
With the \cref{eq_shift_invariance},
 if $0\notin[s_\ii,s_\ff]$, we may apply the shift invariant property to reduce the number of diagrams.
Since each spin has at most $\bar{N}$ couplings,
 the total number of different diagrams with length $l\dt \leqslant L\dt$ is
\begin{equation*}
    \sum_{N=0}^{\bar{N}} (2L+1-l)
    \genfrac(){0pt}{0}{N+l}{N}
    =(2L+1-l) \genfrac(){0pt}{0}{\bar{N}+l+1}{\bar{N}}
\end{equation*}
where the factor $2L+1-l$ is the number of different choices of $s_\ii$, 
 namely, $s_\ii = -L\dt, (-L+1)\dt, \dots, (L-l)\dt$,
 and the binomial coefficient $\genfrac(){0pt}{2}{N+l}{N}$
 represents the different choices of $N$ spin interactions
 on the set $\{s_\ii, s_\ii + \dt,\dots, s_\ii+l\dt\}$.
Practically, when $0\notin[s_\ii,s_\ff]$,
 the translation relation \cref{eq_shift_invariance} can be applied for the reduction of diagrams.
However, the reduction does not change the order of the estimated cost.

Therefore, for the single-spin full propagators of all lengths, the computational cost is estimated by
\begin{equation}
\begin{split}
    & \sum_{l=1}^{2L} (2L+1-l) \genfrac(){0pt}{0}{\bar{N}+l+1}{\bar{N}}
    \sum_{\substack{M=1 \\ M \text{~is odd}}}^{\bar{M}} 
    C_M \genfrac(){0pt}{0}{M+l}{M} \\
    \leqslant {} &
    \sum_{l=1}^{2L} (2L+1-l)
    \genfrac(){0pt}{0}{\bar{N}+l+1}{\bar{N}}
    \frac{C_{\bar{M}}(\bar{M}+1)}{2}
    \genfrac(){0pt}{0}{\bar{M}+l}{\bar{M}} \\
    \lesssim {} & \bar{M}C_{\bar{M}} L
    \sum_{l=1}^{2L} l^{\bar{N}} l^{\bar{M}}
    \lesssim L^{\bar{M}+\bar{N}+2}
\end{split}
\end{equation}
where $\bar{M}, \bar{N}$ are relatively small in practice and are regarded as constants in the above estimation.
For a spin chain with $K$ spins,
 the computational cost should be multiplied by $K$ if all spins have different parameters.

\subsection{Computational cost for the summation}
We now estimate the summation of diagrams described in \cref{sec_resum}.
Note that in this step,
 we only need to use the values of $\mathcal{G}^{(k)}(s_\ii,\boldsymbol{s},s_\ff)$ with $-s_\ii=s_\ff=l\dt$,
 so that the total number of diagrams involved is much less than the previous step.
We now consider the computation of $G^{[k+1]}(-t,\bs,t)$ with $\bs = (s_1, \dots, s_N)$ and $t = l \dt$ according to \cref{add_spin}.
Recall that we have the values of $\mathscr{G}^{(k+1)}(-t,\mathcal{P}(\bs,\bs'), t)$ only for $N + N' \leqslant \bar{N}$ (see the text about the truncation before \Cref{remark_MPI}).
The series \eqref{add_spin} should be truncated up to $N' = \bar{N}$ in the computation.
As a result,
 the computational cost of \cref{add_spin} is
\begin{equation*}
    \sum_{N'=0}^{\bar{N}-N} 
    \genfrac(){0pt}{0}{2l+N'}{N'}
    = \genfrac(){0pt}{0}{1+2l+\bar{N}-N}{\bar{N}-N},
\end{equation*}
where the binomial coefficient on the left-hand side is the number of grid points in the $N'$-dimensional simplex.
Since we need to evaluate $G^{[k+1]}(-t,\bs,t)$ for $\bs$ on all the grid points of an $N$-dimensional simplex,
 and $N$ ranges from $0$ to $\bar{N}$,
 we have the following estimation of the total computational cost:
\begin{equation*}
    \sum_{N=0}^{\bar{N}}
    \genfrac(){0pt}{0}{2l+N}{N}
    \genfrac(){0pt}{0}{1+2l+\bar{N}-N}{\bar{N}-N}
    \lesssim \sum_{N=0}^{\bar{N}}
    l^{N} l^{\bar{N}-N}
    \lesssim \bar{N} l^{\bar{N}}.
\end{equation*}
Finally,
 to compute observables on all time steps $l = 1,\dots,L$,
 the time complexity is then $\mathcal{O}(L^{\bar{N}+1})$.
Compared to the solver of the inchworm equation,
 the computational cost of the summation is relatively small.
Hence, the total computational cost remains at $\mathcal{O}(L^{\bar{M}+\bar{N}+2})$
 as analyzed in \cref{sec_computational_cost_one_spin}.

\subsection{Numerical verification}
\begin{figure}
    \centering
    \includegraphics[width=0.4\textwidth]{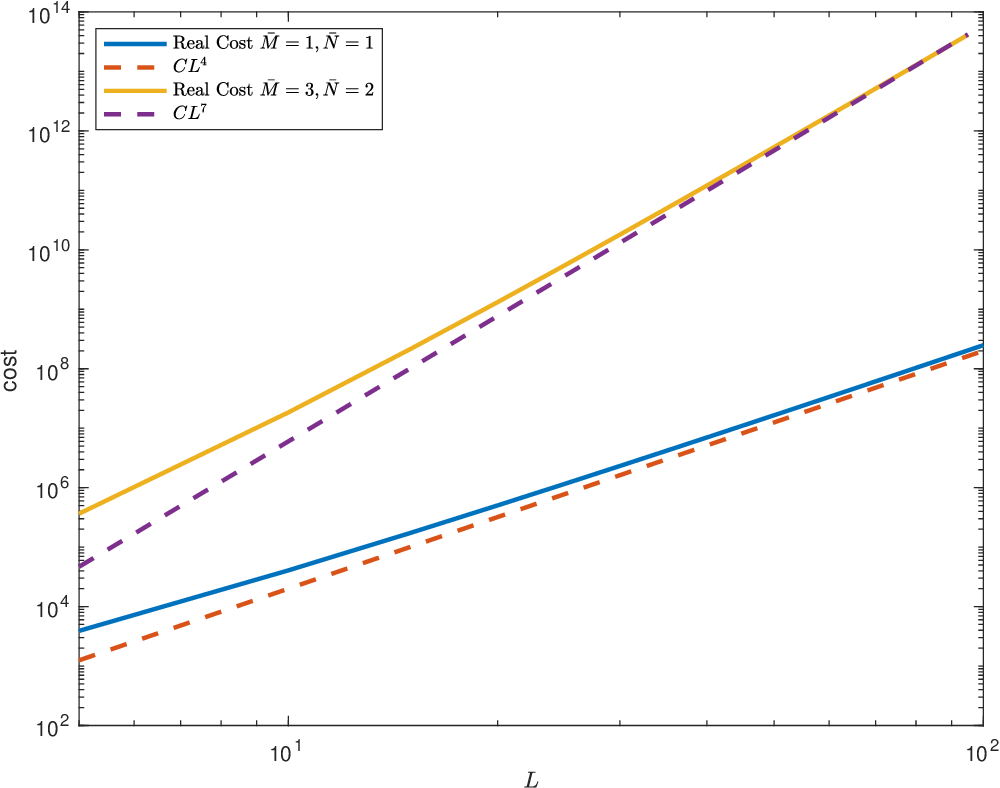}
    \caption{Computational cost for different $L$ ($\bar{M}=1,\bar{N}=1$ and $\bar{M}=3,\bar{N}=2$).}
    \label{fig_computational_cost}
\end{figure}

In agreement with our analysis,
 our numerical experiments (to be presented in detail in \cref{sec_numerical}) also show that the computational cost of the summation is nearly negligible compared with the solver of the inchworm equation.
Therefore, to verify our estimation of the computational cost,
 we will focus only on the analysis in \cref{sec_computational_cost_one_spin}.
A convenient way to check the time complexity is to count the number of evaluations of the bath influence functional $\mathcal{L}_b^{(c)}(\bs)$, which depends only on $L, \bar{M}, \bar{N}$ and is independent of all other parameters.
Results for $\bar{M} = 1, \bar{N} = 1$ and $\bar{M} = 3, \bar{N} = 2$ with different values of $L$ are plotted in \cref{fig_computational_cost}.
It can be clearly seen that when $L$ gets larger, the trend of growth agrees better with our analysis.

In general,
 this estimation of the computational cost is the same as direct path-integral methods such as the summation of the Dyson series.
However,
 the use of bold lines can significantly accelerate the convergence of the series,
 resulting in a much smaller $\bar{M}$ needed in the simulation.
The time complexity $\mathcal{O}(L^{\bar{M} + \bar{N} + 2})$ shows that reducing $\bar{M}$ has a great impact on the computational cost,
 especially for large values of $L$.

\begin{remark}
We would like to comment that in the algorithm,
 the most time-consuming step is the evaluation of $\mathscr{G}^{(k)}{(s_\ii, \bs, s_\ff)}$.
To reduce the computational time,
 multithreading is implemented to parallelize the computation.
In general,
 according to the structure of the inchworm equation \eqref{inchworm_differential_equation},
 the value of $\mathscr{G}^{(k)}{(s_\ii, \bs, s_\ff)}$ for shorter $\bs$ is needed to obtain the full propagator for longer $\bs$.
Therefore, we first compute $\mathscr{G}^{(k)}{(s_\ii, \varnothing, s_\ff)}$ for all $s_\ii,s_\ff$,
 and the solve $\mathscr{G}^{(k)}{(s_\ii, s_1, s_\ff)}$ for all $s_\ii, s_1, s_\ff$, followed by the computation of $\mathscr{G}^{(k)}{(s_\ii, s_1,s_2, s_\ff)}$ for all $s_\ii, s_1, s_2, s_\ff$, and so forth until the maximum length of $\bs$ is reached.
The computations of $\mathscr{G}^{(k)}{(s_\ii, \varnothing, s_\ff)}$ and $\mathscr{G}^{(k)}{(s_\ii, s_1, s_\ff)}$ are carried out sequentially.
When the length of $\bs$ in $\mathscr{G}^{(k)}{(s_\ii, \bs, s_\ff)}$ is greater than or equal to 2,
 the algorithm is parallelized.
The parallelization is based on the fact that the inchworm equations \eqref{inchworm_differential_equation} for $\mathscr{G}^{(k)}{(s_\ii, s_1, \dots, s_N, s_\ff)}$ can actually be decoupled.
Precisely speaking,
 the propagator $\mathscr{G}^{(k)}{(s_\ii', s_1', \dots, s_N', s_\ff')}$ can appear on the right-hand side of \cref{inchworm_differential_equation} for $\bs = (s_1, \dots, s_N)$ only when $s_k' = s_k$ for all $k = 1,\dots,N$,
 and in this case,
 we have $s_\ii' = \tau_m$ and $s_\ff' = \tau_{m+1}$ for a certain $m$.
If $0 \not\in (s_\ii', s_\ff')$,
 the value of $\mathscr{G}^{(k)}{(s_\ii', s_1', \dots, s_N', s_\ff')}$ (or $\mathscr{G}^{(k)}{(\tau_m, s_1, \dots, s_N, \tau_{m+1})}$) is actually obtained from \cref{eq_shift_invariance}.
Therefore, $\mathscr{G}^{(k)}{(s_\ii, s_1, \dots, s_N, s_\ff)}$ and $\mathscr{G}^{(k)}{(s_\ii', s_1', \dots, s_N', s_\ff')}$ are coupled only if there exists $T$ such that $s_j' = s_j + T$ for all $j = 1,\cdots,N$.
This allows decoupling of equations according to the vector $(s_2 - s_1, \dots, s_N - s_{N-1})$, and thus the algorithm can be parallelized.

In fact, when $0 \in (s_1, s_N)$,
 the equations of $\mathscr{G}^{(k)}(s_\ii, \bs, s_\ff)$ are decoupled simply for different values of $\bs$, since the translational relation \cref{eq_shift_invariance} cannot be applied. Using this structure helps with better distribution of computational cost across the threads.
\end{remark}

\section{Numerical Experiments}
\label{sec_numerical}
In this section, we evaluate our newly-proposed method using several numerical examples. 
To begin with, we introduce the parameters used for the numerical tests.
For the coupling intensity between spins, the operator $V^{(k)}$ simply a scaled Pauli matrix:
\begin{equation}
    V^{(k)} = J^{(k)} \sigma_z^{(k)}
\end{equation}
where $J^{(k)}$ indicates the coupling intensity between the $k$th spin and its neighboring spins.
The observable is chosen to be $O_s = \sigma_z^{(k)}$ for $k=1,\dots,K$, respectively.
In \cref{eq_mathscrGkdefn}, the two point correlation functions $B^{(k)}(\tau_1,\tau_2)$ are set to be the same for every $k$:
\begin{equation}
\label{bath_function_B}
    B^{(k)}(\tau_j,\tau_{j'}) = B^*(\Delta\tau)
    =\frac{1}{\pi} \int_0^{\infty}
    J(\omega)
    \left[
    \coth\left(\frac{\beta\omega}{2}\right) \cos(\omega\Delta t)
    - \ii \sin(\omega \Delta t)
    \right] \dd \omega
\end{equation}
where $\Delta \tau = \vert \tau_j \vert - \vert \tau_{j'} \vert$ and $J(\omega)$ is the spectral density of the harmonic oscillators in the bath.
In this paper, we set it to be the Ohmic spectral density:
\begin{equation}
    J(\omega) = \frac{\pi}{2} \sum_{l=1}^L \frac{c_l^2}{\omega_l} \delta(\omega - \omega_l) 
\end{equation}
where $L$ is the number of harmonic oscillators and is set to be $400$ in all our tests.
The coupling intensity $c_l$ and frequency of each harmonic oscillator $\omega_l$ are given by
\begin{align*}
    &\omega_l = -\omega_c \ln \left(1- \frac{l}{L}[1-\exp(-\omega_{\max}/\omega_c)]\right), \\
    & c_l = \omega_l \sqrt{ \frac{\xi \omega_c}{L}[1-\exp(-\omega_{\max}/\omega_c)]}.
\end{align*}
The values of the parameters, including the Kondo parameter $\xi$, the primary frequency of the harmonic oscillators $\omega_c$, and the maximum frequency $\omega_{\max}$, will be given later for each experiment.

In addition to the above physical parameters,
 three numerical parameters need to be specified to carry out the simulation,
 including two truncation parameters ($\bar{M}$ for system-bath couplings and $\bar{N}$ for interspin couplings) and the time step $\dt$.
The convergence of the numerical results with respect to these parameters will be studied in the following subsection.

\subsection{Convergence tests}
This section carries out experiments on three convergence parameters, $\bar{M},\bar{N}$ and $\dt$,
among which $\bar{M},\bar{N}$ are two truncation parameters and $\dt$ stands for the time step.
In this section, all spins in the spin chain are prepared in the state $\ket{+1}$.
In other words, $\varsigma^{(k)} = +1$ for $k=1,\dots,K$ in \cref{eq_Initial_condition_general}.

In the spin-boson model with a single spin,
 the convergence with respect to the parameter $\bar{M}$ has been studied numerically in \cite{chen2017inchworm2, cai2020inchworm},
 where it was shown that the convergence of the inchworm method was much faster than the Dyson series.
Here we will carry out a numerical test for the convergence of $\bar{M}$ by considering a 5-spin system. We choose the time step to be $\dt = 0.2$.
Other parameters are chosen as follows:
\begin{gather*}
    \xi = 0.2,
    \quad \beta = 5,
    \quad \omega_c = 2.5,
    \quad \omega_{\max} = 4\omega_c, \quad \bar{N} = 2, \\
    \epsilon^{(k)} = 1,
    \quad \Delta^{(k)} = 1,
    \quad J^{(k)} = 0.2, 
    \qquad \forall k = 1,\dots,5.
\end{gather*}
\begin{figure}[t]
    \centering
    \begin{subfigure}[b]{0.3\textwidth}
         \centering
         \includegraphics[width=\textwidth]{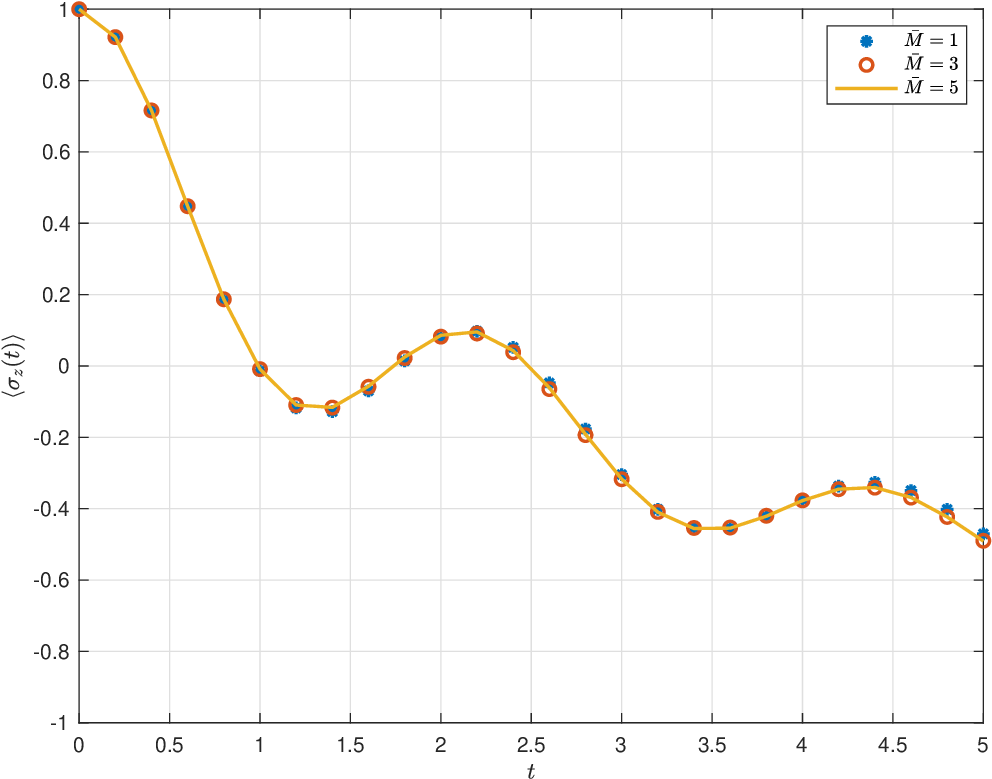}
         \caption{Spins 1 and 5}
         \label{fig_convergence_M_spin1}
     \end{subfigure}
     \hfill
    \begin{subfigure}[b]{0.3\textwidth}
         \centering
         \includegraphics[width=\textwidth]{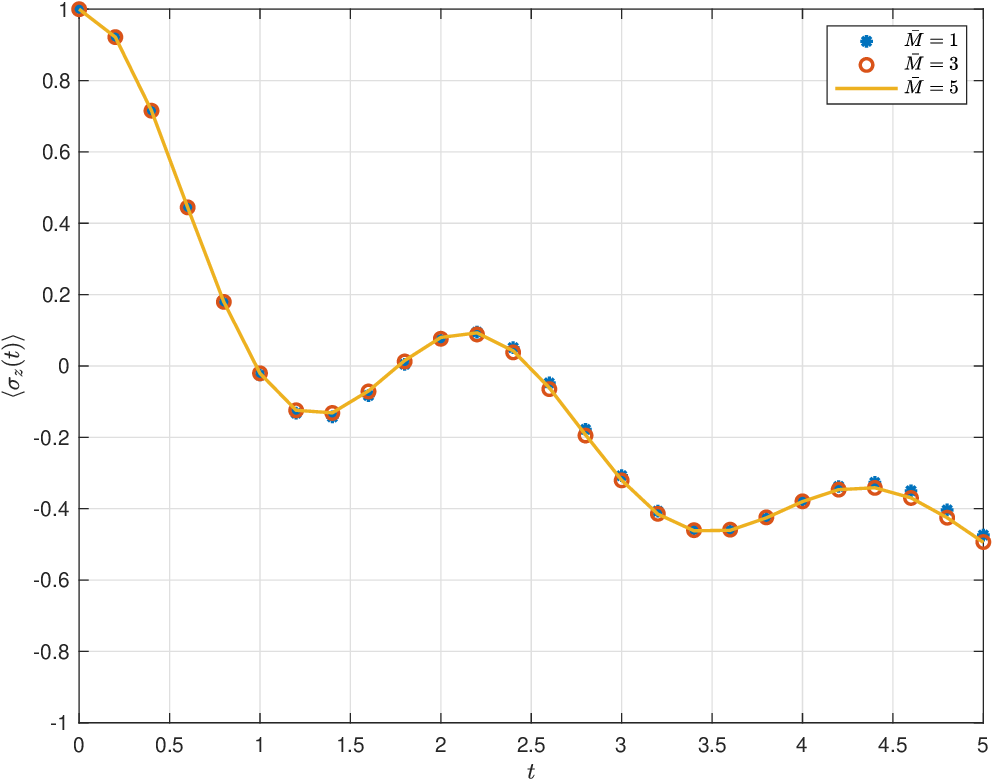}
         \caption{Spins 2 and 4}
         \label{fig_convergence_M_spin2}
     \end{subfigure}
     \hfill
     \begin{subfigure}[b]{0.3\textwidth}
         \centering
         \includegraphics[width=\textwidth]{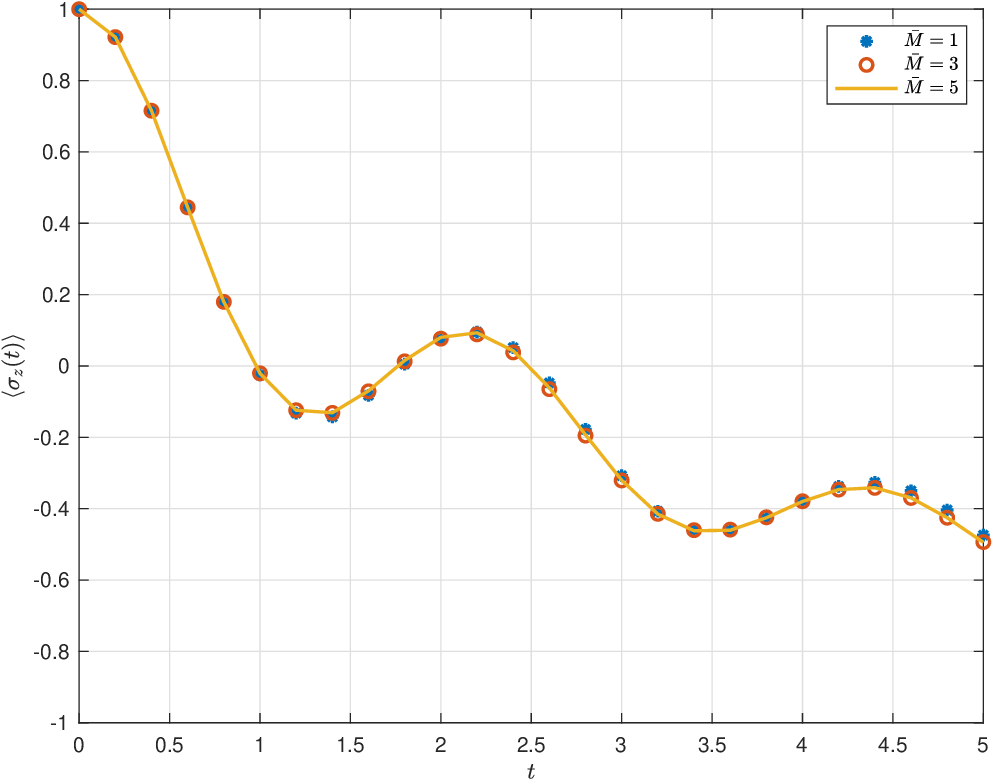}
         \caption{Spin 3}
         \label{fig_convergence_M_spin3}
     \end{subfigure}
    \caption{Convergence of the method with respect to $\bar{M}$.}
    \label{fig_convergence_M}
\end{figure}%
Our numerical results are given in \cref{fig_convergence_M},
 which shows the evolution of $\langle \sigma_z^{(k)} \rangle$ for $k = 1,\cdots,5$.
Note that due to the symmetry of the spin chain system,
 we have $\expval{\sigma_z^{(1)}(t)} = \expval{\sigma_z^{(5)}(t)}$ and $\expval{\sigma_z^{(2)}(t)} = \expval{\sigma_z^{(4)}(t)}$ for all $t$,
 and therefore only three figures are shown in \cref{fig_convergence_M}.
These figures show fast convergence with respect $\bar{M}$ for this set of parameters,
 due to the use of the inchworm method.
The curves for $\bar{M} = 3$ and $\bar{M} = 5$ are almost on top of each other,
 while some slight differences can be observed for the computation with $\bar{M} = 1$,
 which is less accurate.

\begin{figure}
    \centering
    \begin{subfigure}[b]{0.3\textwidth}
         \centering
         \includegraphics[width=\textwidth]{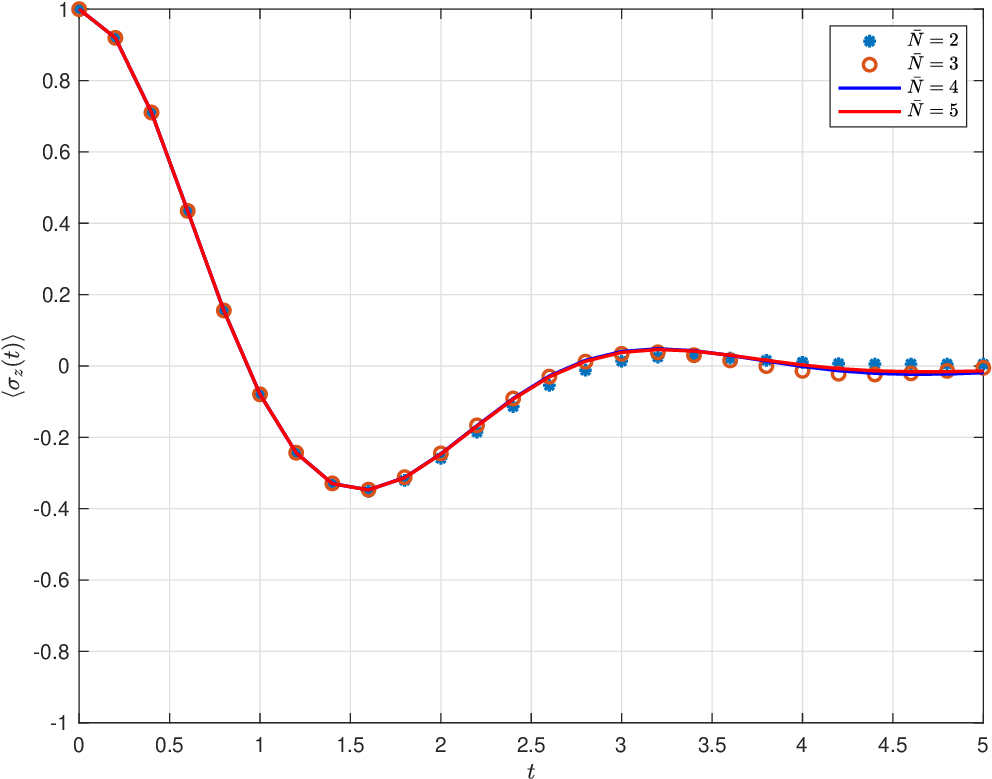}
         \caption{Spins 1 and 5}
         \label{fig_convergence_spin1}
     \end{subfigure}
     \hfill
    \begin{subfigure}[b]{0.3\textwidth}
         \centering
         \includegraphics[width=\textwidth]{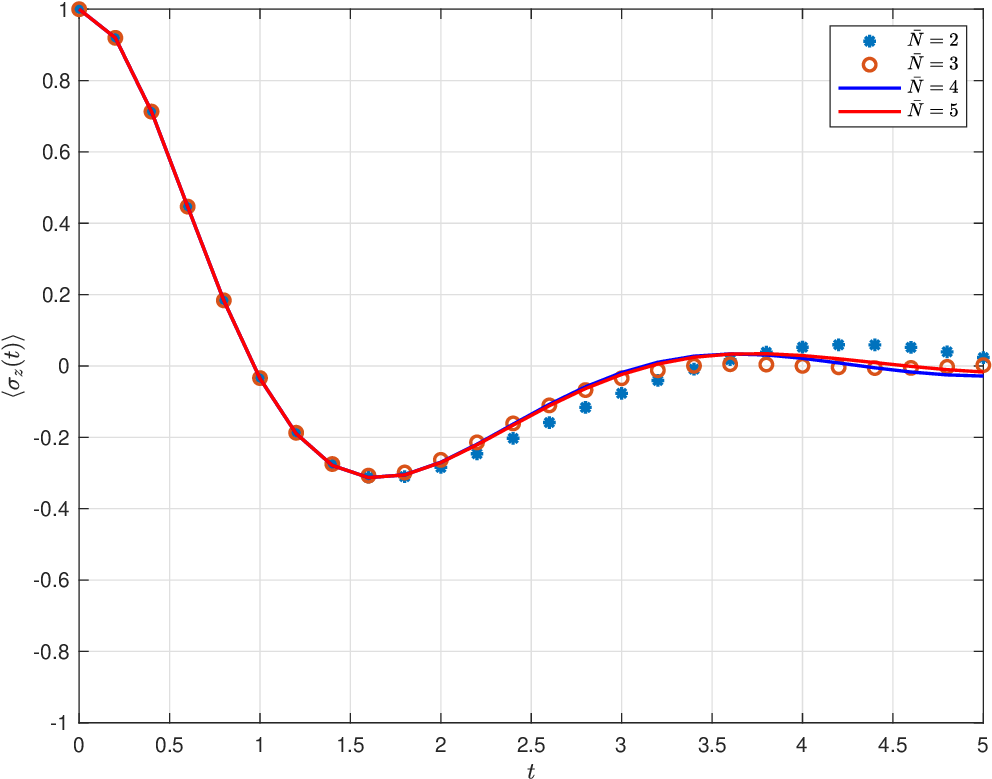}
         \caption{Spins 2 and 4}
         \label{fig_convergence_spin2}
     \end{subfigure}
     \hfill
     \begin{subfigure}[b]{0.3\textwidth}
         \centering
         \includegraphics[width=\textwidth]{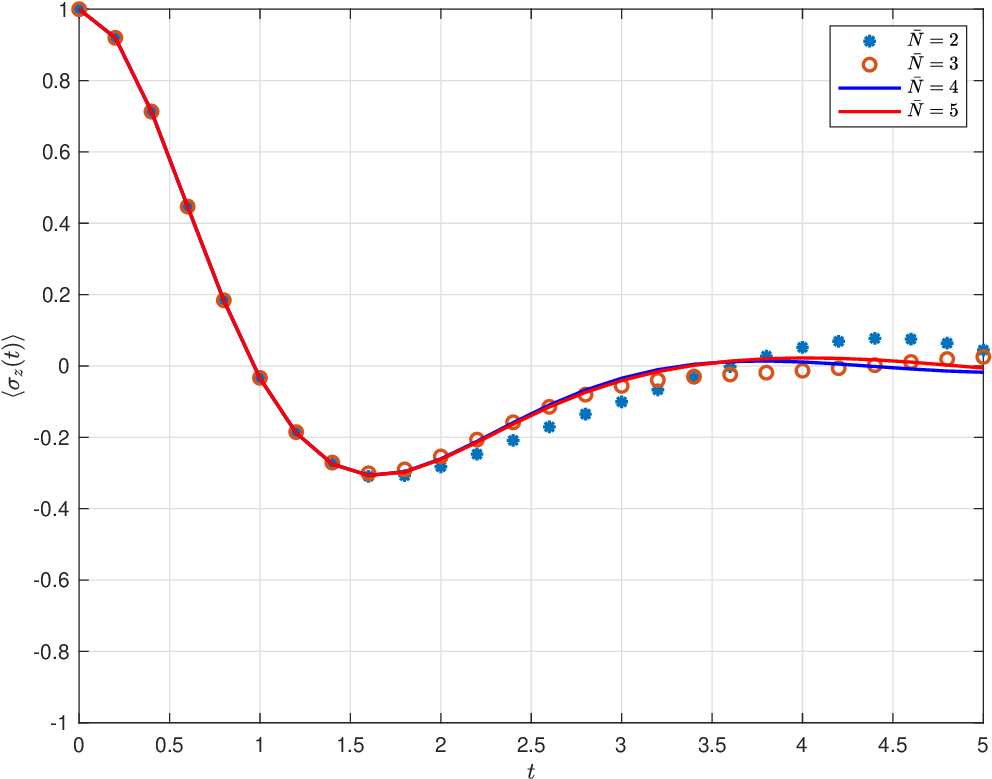}
         \caption{Spin 3}
         \label{fig_convergence_spin3}
     \end{subfigure}
    \caption{Convergence of the method with respect to $\bar{N}$.}
    \label{fig_convergence_N}
\end{figure}

We now fix $\bar{M}$ and consider the convergence with respect to $\bar{N}$.
We again consider a chain of 5 spins 
and choose the time step to be $\dt = 0.2$. 
Other parameters are
\begin{gather*}
    \xi = 0.2,
    \quad \beta = 5,
    \quad \omega_c = 2.5,
    \quad \omega_{\max} = 4\omega_c, \quad \bar{M} = 3, \\
    \epsilon^{(k)} = 0,
    \quad \Delta^{(k)} = 1,
    \quad J^{(k)} = 0.5, 
    \qquad \forall k = 1,\dots,5.
\end{gather*}
The results for $\bar{N}=2,3,4,5$ are shown in \cref{fig_convergence_N}.
In general,
 due to the numerical sign problem,
 for longer-time simulations,
 larger values of $\bar{N}$ are needed to obtain accurate results.
For the first and the last spins,
 since they are coupled only with one neighboring spin,
 the results of $\bar{N} = 3$ already show good quality until $t = 5$.
For the remaining three spins,
 the results for $\bar{N} = 4$ and $\bar{N} = 5$ almost coincide,
 showing the convergence for the coupling intensity $J^{(k)} = 0.5$ up to $t = 5$.
Further increasing $\bar{N}$ does not significantly improve the results. 

\begin{figure}
    \centering
    \begin{subfigure}[b]{0.3\textwidth}
         \centering
         \includegraphics[width=\textwidth]{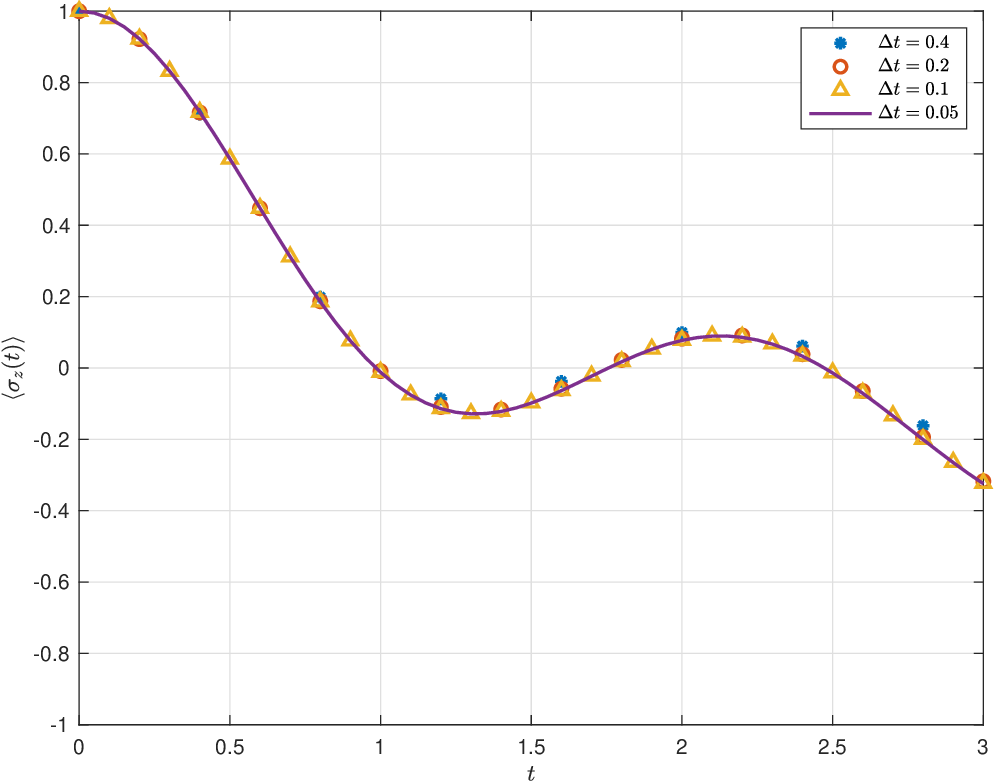}
         \caption{Spins 1 and 5}
         \label{fig_convergence_dt_spin1}
     \end{subfigure}
     \hfill
    \begin{subfigure}[b]{0.3\textwidth}
         \centering
         \includegraphics[width=\textwidth]{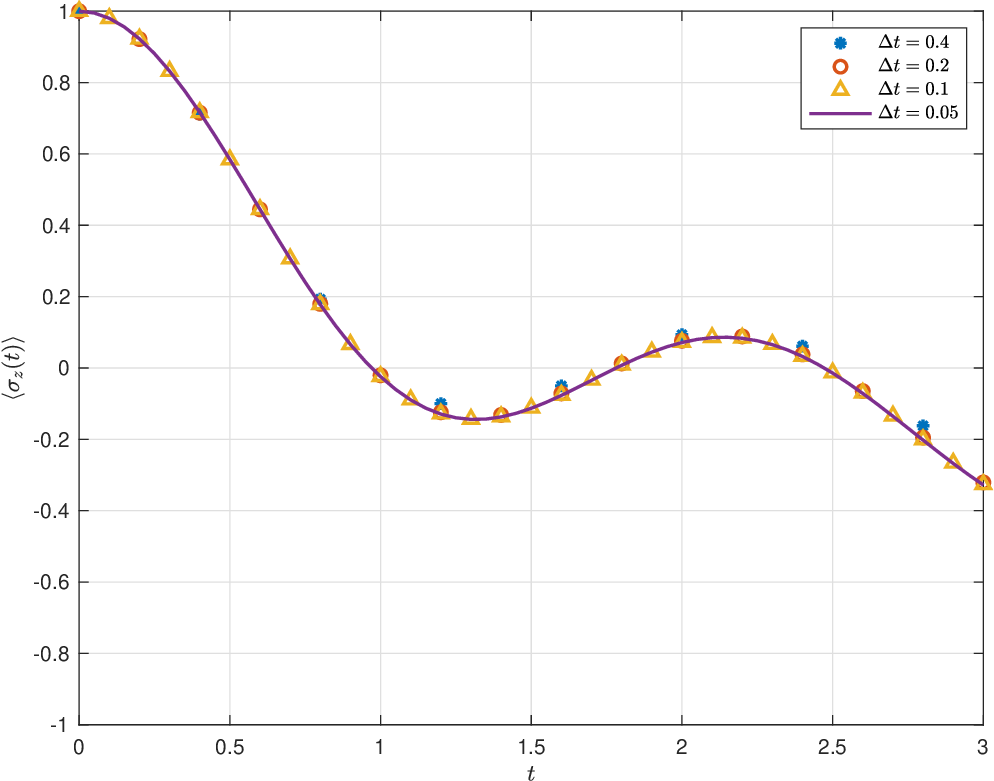}
         \caption{Spins 2 and 4}
         \label{fig_convergence_dt_spin2}
     \end{subfigure}
     \hfill
     \begin{subfigure}[b]{0.3\textwidth}
         \centering
         \includegraphics[width=\textwidth]{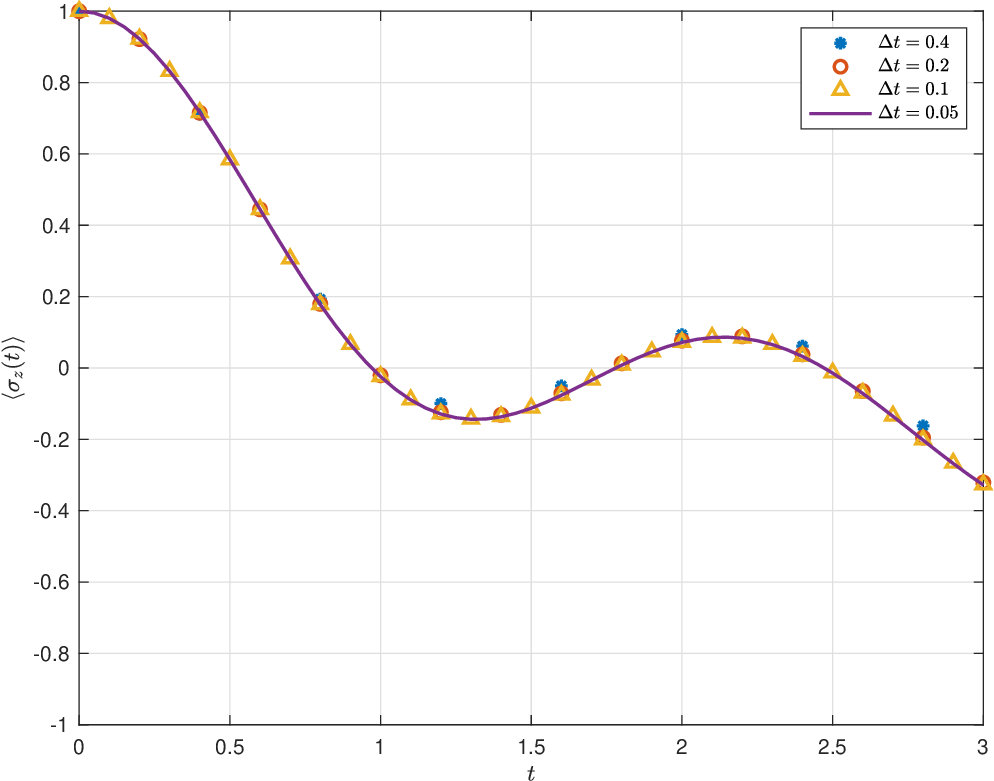}
         \caption{Spin 3}
         \label{fig_convergence_dt_spin3}
     \end{subfigure}
    \caption{Convergence of the method with respect to $\dt$.}
    \label{fig_convergence_dt}
\end{figure}

Additionally, the convergence test is also carried out for the time step $\dt$, with the parameters of the $5$-spin Ising chain being
\begin{gather*}
    \xi = 0.2,
    \quad \beta = 5,
    \quad \omega_c = 2.5,
    \quad \omega_{\max} = 4\omega_c, 
    \quad \bar{M} = 3, 
    \quad \bar{N} = 2\\
    \epsilon^{(k)} = 1,
    \quad \Delta^{(k)} = 1,
    \quad J^{(k)} = 0.2, 
    \qquad \forall k = 1,\dots,5.
\end{gather*}
We perform simulations for the time step $\dt$ being $0.4$, $0.2$, $0.1$, and $0.05$ and present the results in \cref{fig_convergence_dt}.
Note that for $\bar{M} = 3$ and $\bar{N} = 2$,
 according to our analysis in \cref{sec_computational_cost},
 the computational cost is estimated by $\mathcal{O}(L^7)$ with $L$ being the total number of time steps.
Therefore, to save computational time,
 we run the simulation only up to $t = 3$.
It can be observed that for our second-order numerical method,
 the time step $\Delta t = 0.2$ can give sufficiently accurate results.
Such a time step will be taken for all the simulations in the following subsections.

\subsection{Numerical tests for different coupling intensities}

\begin{figure}
    \centering
    \begin{subfigure}[b]{0.4\textwidth}
         \centering
         \includegraphics[width=\textwidth]{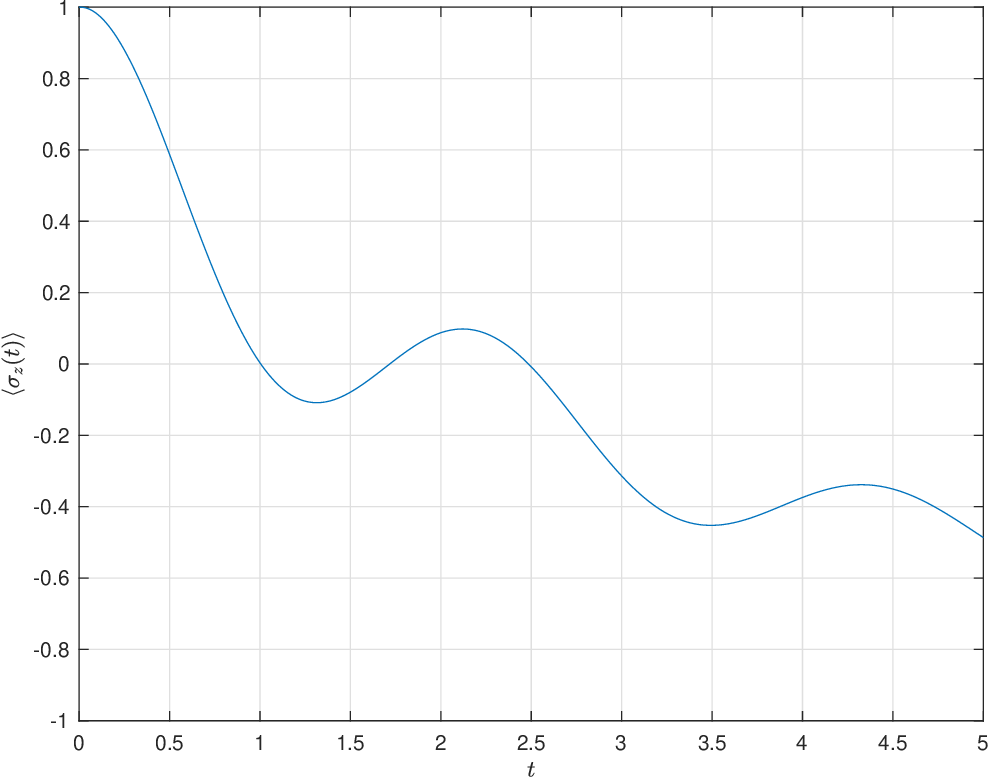}
         \caption{Single spin (coupling intensity $J^{(k)}=0$)}
         \label{fig_different_coupling_00}
    \end{subfigure}
    \begin{subfigure}[b]{0.4\textwidth}
         \centering
         \includegraphics[width=\textwidth]{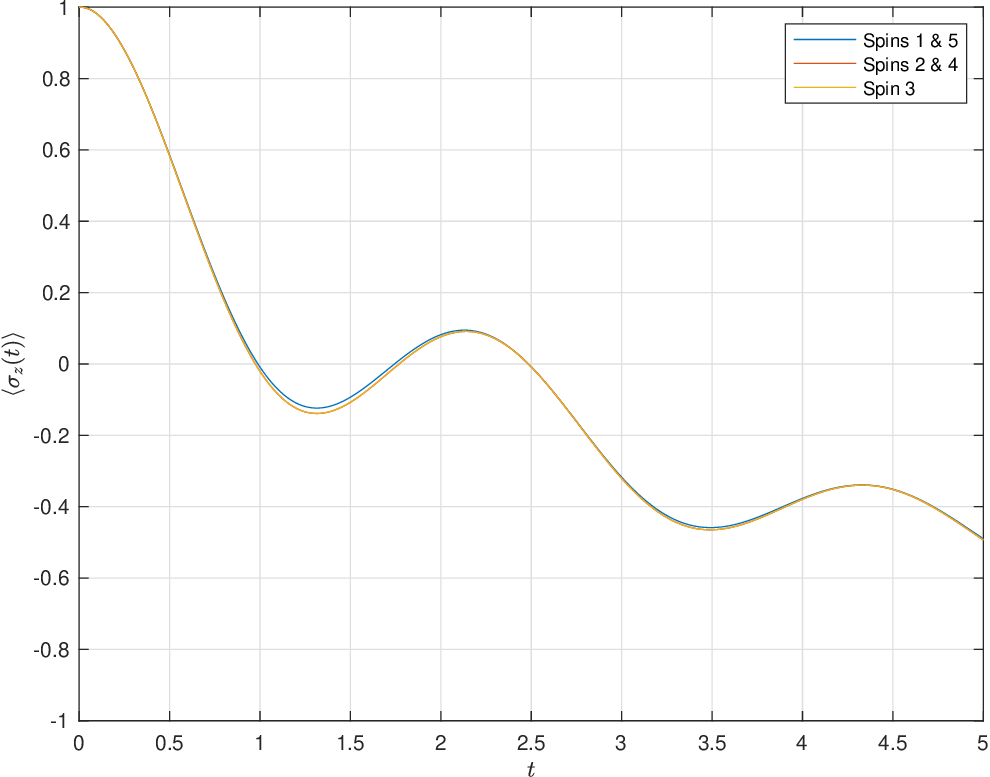}
         \caption{Coupling intensity $J^{(k)}=0.2$.}
         \label{fig_different_coupling_02}
     \end{subfigure}
     \begin{subfigure}[b]{0.4\textwidth}
         \centering
         \includegraphics[width=\textwidth]{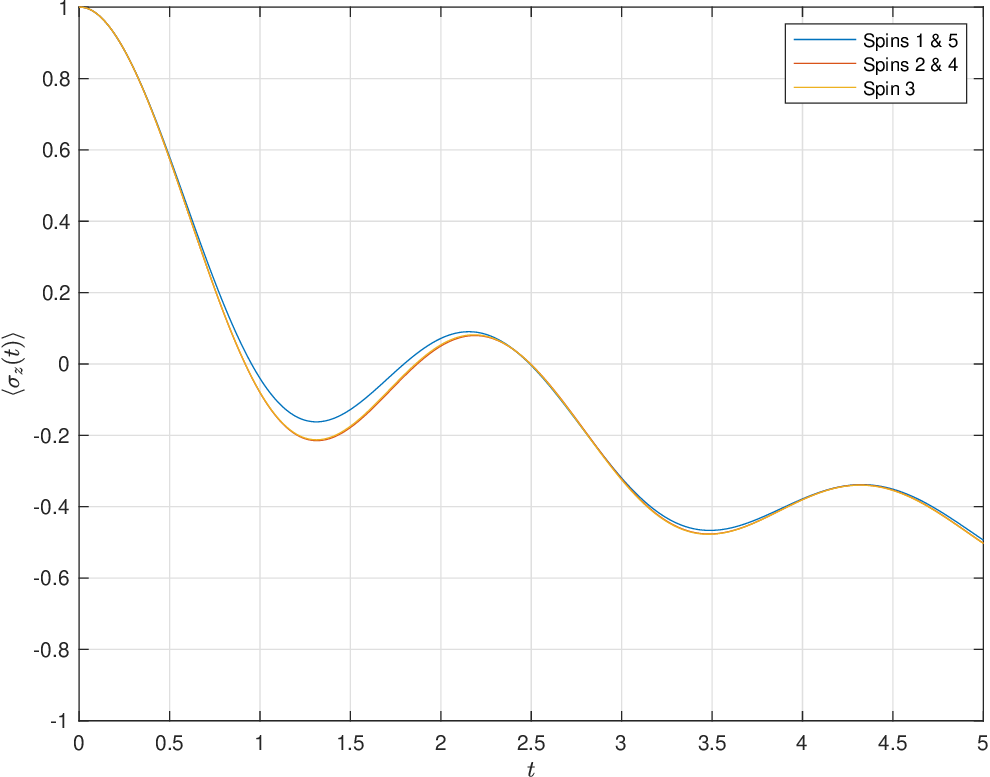}
         \caption{Coupling intensity $J^{(k)}=0.4$.}
         \label{fig_different_coupling_04}
     \end{subfigure}
     \begin{subfigure}[b]{0.4\textwidth}
         \centering
         \includegraphics[width=\textwidth]{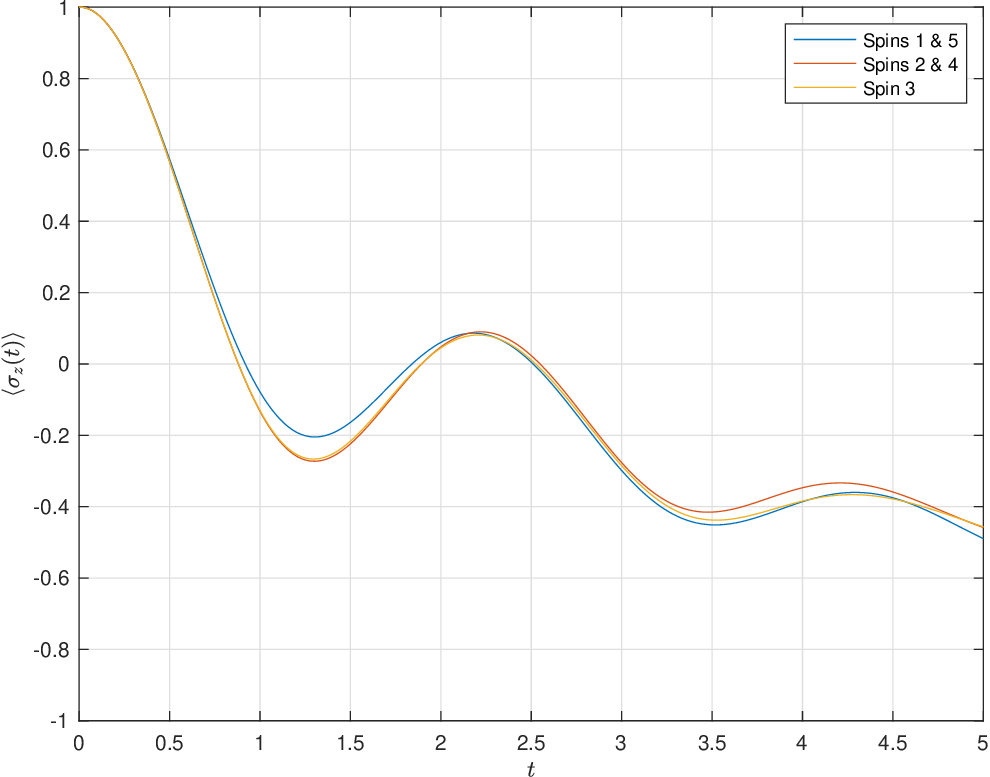}
         \caption{Coupling intensity $J^{(k)}=0.6$.}
         \label{fig_different_coupling_06}
     \end{subfigure}
    \caption{Numerical Experiments for different coupling intensity between spins.}
    \label{fig_result_different_coupling_intensity}
\end{figure}

In this section, we conduct numerical experiments to examine the effects of varying coupling intensities between spins.
We again consider the 5-spin Ising chain with the following parameters:
\begin{gather*}
    \xi = 0.2,
    \quad \beta = 5,
    \quad \omega_c = 2.5,
    \quad \omega_{\max} = 4\omega_c, 
    \quad \bar{M} = 3, 
    \quad \bar{N} = 4\\
    \epsilon^{(k)} = 1,
    \quad \Delta^{(k)} = 1,
    \qquad \forall k = 1,\dots,5.
\end{gather*}
As mentioned previously,
 the time step is chosen as $\Delta t = 0.2$, which is sufficient to guarantee a small truncation error.
We again set $J^{(k)}$ to be the same for all $k = 1,\dots,5$.
Three values $J^{(k)} = 0.2$, $0.4$, $0.6$ are considered in our experiments,
 and the results are given in \cref{fig_result_different_coupling_intensity}
 given that all spins are initially in the state $\ket{\varsigma^{(k)}} = \ket{+1}$ for all $k$.
Again, our results correctly reflect the symmetry of the Ising chain,
 and therefore only three lines are plotted in each figure.

For the purpose of comparison, we also include the result for $J^{(k)} = 0$,
 meaning that all the spins are decoupled.
In this case,
 the evolution of the observable is identical for all the spins,
 and they are the same as the spin-boson model studied in \cite{cai2020inchworm}.

Generally, for higher coupling intensity $J^{(k)}$,
 the discrepancy between spins is more significant,
 and they differ more from the decoupled case.
In particular,
 when $J^{(k)} = 0$,
 all the curves coincide as predicted.
It can also be observed that the curve for the first and the last spin is more separated from the other three spins,
 especially in the initial stage of the dynamics.
This is due to the fact that the two spins at the ends of the chain interact only with one spin instead of two.
In all cases,
 the interaction between the spin and the bath causes smaller amplitude of the fluctuation as the system evolves.

\begin{figure}
    \centering
    \begin{subfigure}[b]{0.4\textwidth}
         \centering
         \includegraphics[width=\textwidth]{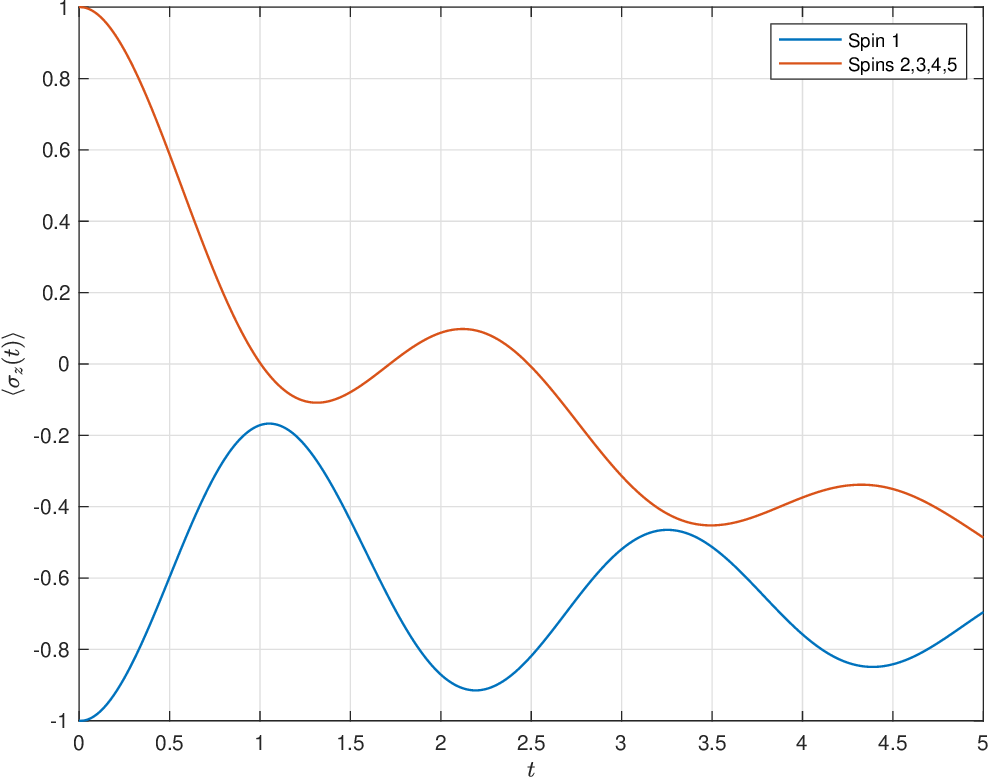}
         \caption{Decoupled spins ($J^{(k)}=0$)}
         \label{fig_different_coupling_00_spin1_down}
    \end{subfigure}
    \begin{subfigure}[b]{0.4\textwidth}
         \centering
         \includegraphics[width=\textwidth]{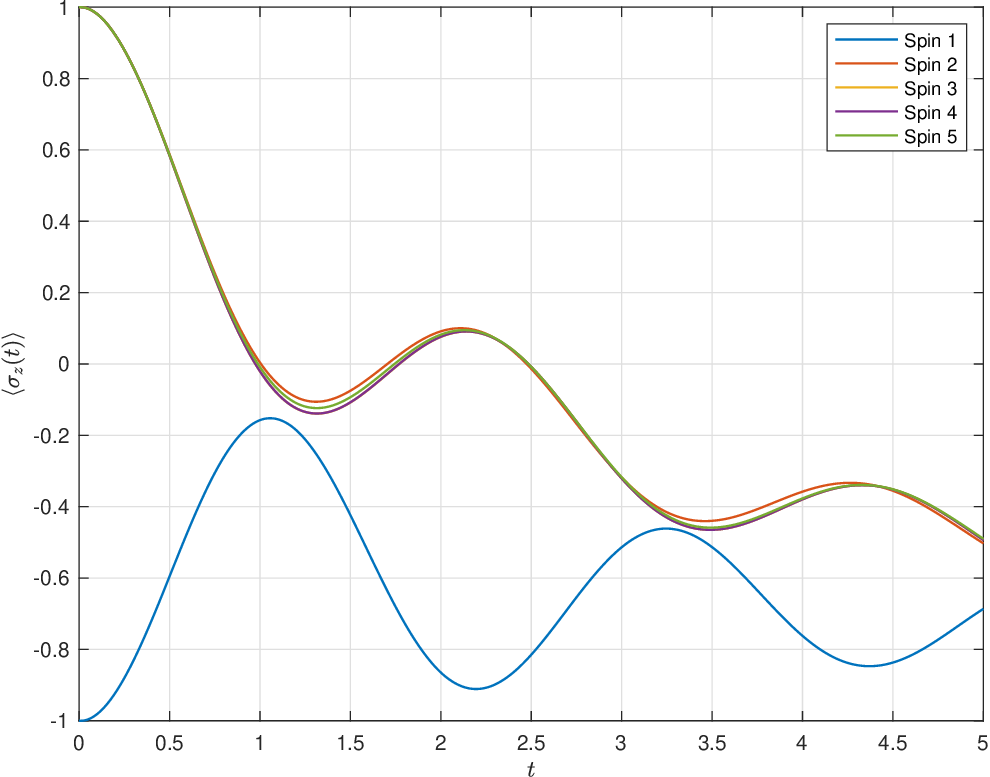}
         \caption{Coupling intensity $J^{(k)}=0.2$.}
         \label{fig_different_coupling_02_spin1_down}
     \end{subfigure}
     \begin{subfigure}[b]{0.4\textwidth}
         \centering
         \includegraphics[width=\textwidth]{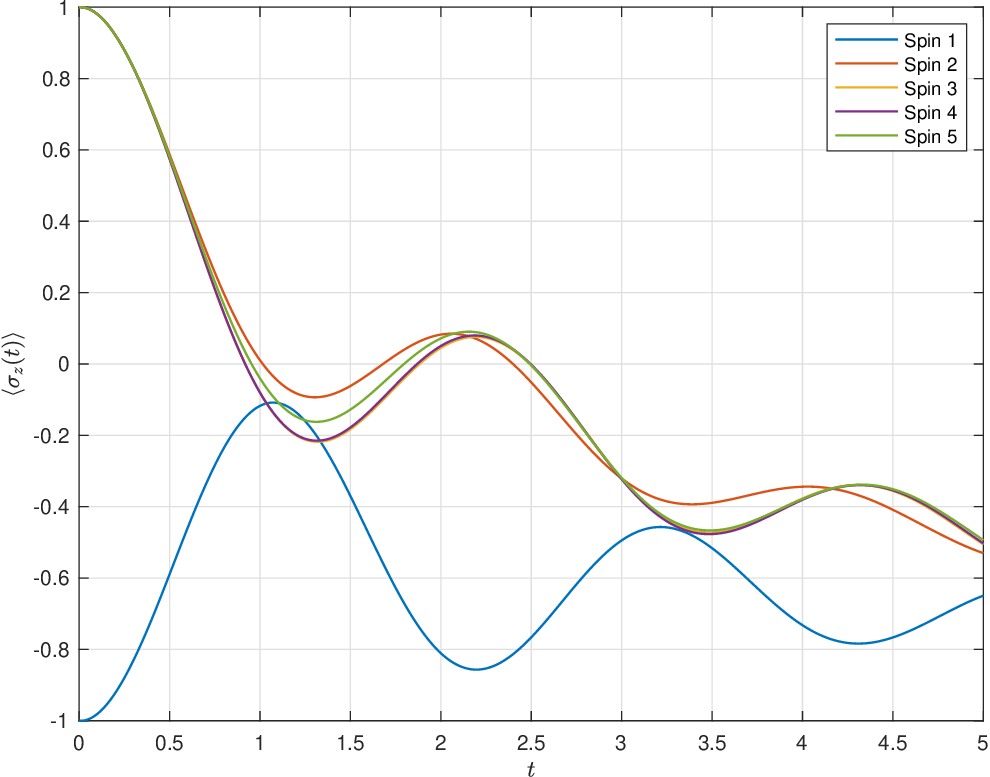}
         \caption{Coupling intensity $J^{(k)}=0.4$.}
         \label{fig_different_coupling_04_spin1_down}
     \end{subfigure}
     \begin{subfigure}[b]{0.4\textwidth}
         \centering
         \includegraphics[width=\textwidth]{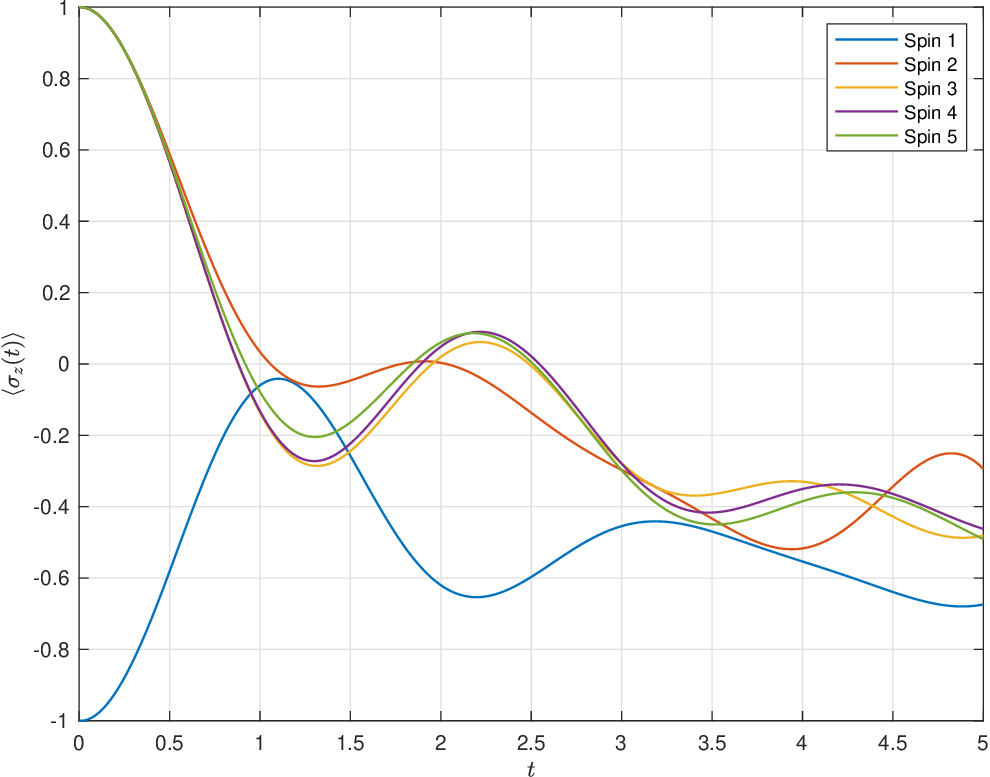}
         \caption{Coupling intensity $J^{(k)}=0.6$.}
         \label{fig_different_coupling_06_spin1_down}
     \end{subfigure}
    \caption{Numerical Experiments for different coupling intensity between spins (Spin 1 down initially).}
    \label{fig_result_different_coupling_intensity_spin1_down}
\end{figure}

Additionally, we also carry out an experiment where the first spin is initially at the state $\ket{\varsigma^{(1)}} = \ket{-1}$ and all other spins have the initial state $\ket{\varsigma^{(k)}} = \ket{+1}$ for $k=2,\dots,5$.
Such a spin chain is no longer symmetric.
The evolution of the observable $\langle \sigma_z^{(k)}(t) \rangle$ is plot in \cref{fig_result_different_coupling_intensity_spin1_down}.
In this experiment, when $J^{(k)}=0$,
 Spins 2 to 5 are physically identical,
 so there are only two distinct curves in the figure.
For non-zero coupling intensities between spins,
 it is clear that the behavior of the first spin is affected by the other spins.
The local minimum of the blue curves around $t = 2.2$ is obviously higher when the coupling intensity $J^{(k)}$ gets larger.
Similar to \cref{fig_result_different_coupling_intensity},
 the separation of the curves for Spins 2 to 5 also gets clearer for stronger coupling between spins.

\subsection{Simulation of a long Ising chain}

This section aims to study the behavior of a long spin chain, in which the middle part can mimic the behavior of an infinite Ising chain,
 and meanwhile, one can observe the end effects.
We consider an Ising chain comprising of 50 spins and 100 spins, respectively.
The parameters of all the spins are set to be the same.
Under such settings,
 we anticipate observing very similar behaviours for the spins near the center of the chain.
Note that in our method, if the spins and baths have the same physical parameters,
 the computational cost grows only linearly as the number of spins increases.
The parameters used in this experiment are
\begin{gather*}
    \xi = 0.2,
    \quad \beta = 5,
    \quad \omega_c = 2.5,
    \quad \omega_{\max} = 4\omega_c, 
    \quad \bar{M} = 3, 
    \quad \bar{N} = 4\\
    \epsilon^{(k)} = 0,
    \quad \Delta^{(k)} = 1,
    \quad J^{(k)} = 0.5,
    \qquad \forall k = 1,\dots,K.
\end{gather*}
with $K=50$ or $K=100$.
The time step is chosen as $\dt = 0.2$.
For comparison, 
 we also carry out the experiments for the same parameters with $K=1$ and $K=5$.

\begin{figure}
    \centering
    \begin{subfigure}[b]{0.4\textwidth}
         \centering
         \includegraphics[width=\textwidth]{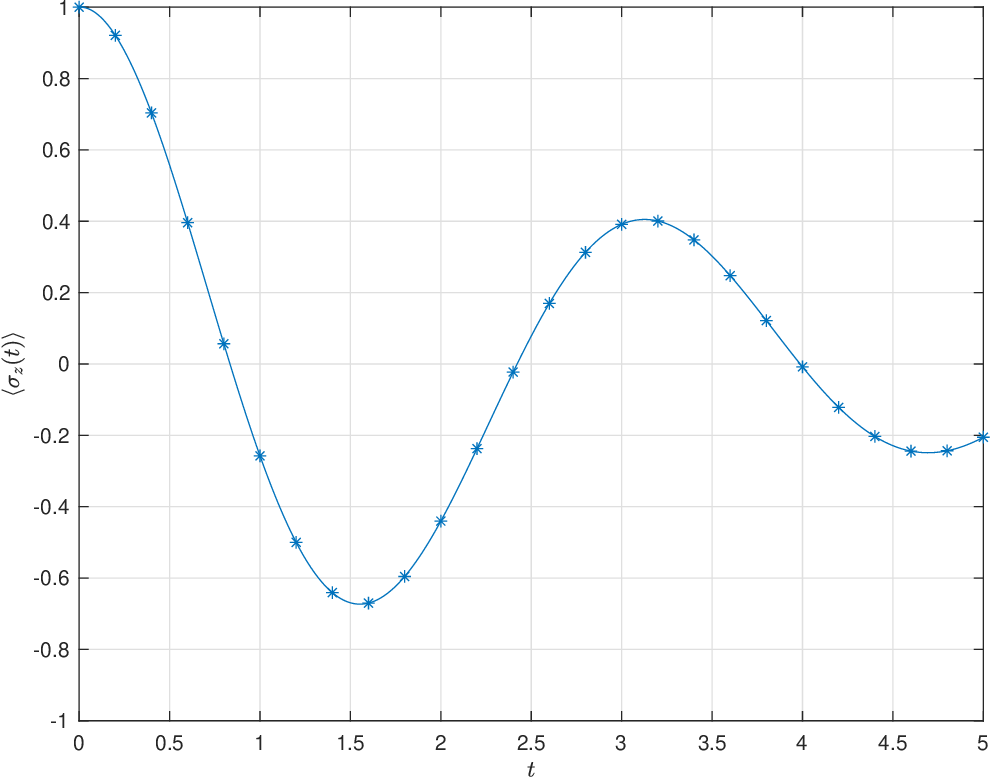}
         \caption{Single spin.}
         \label{fig_1_spin}
    \end{subfigure}
    \begin{subfigure}[b]{0.4\textwidth}
         \centering
         \includegraphics[width=\textwidth]{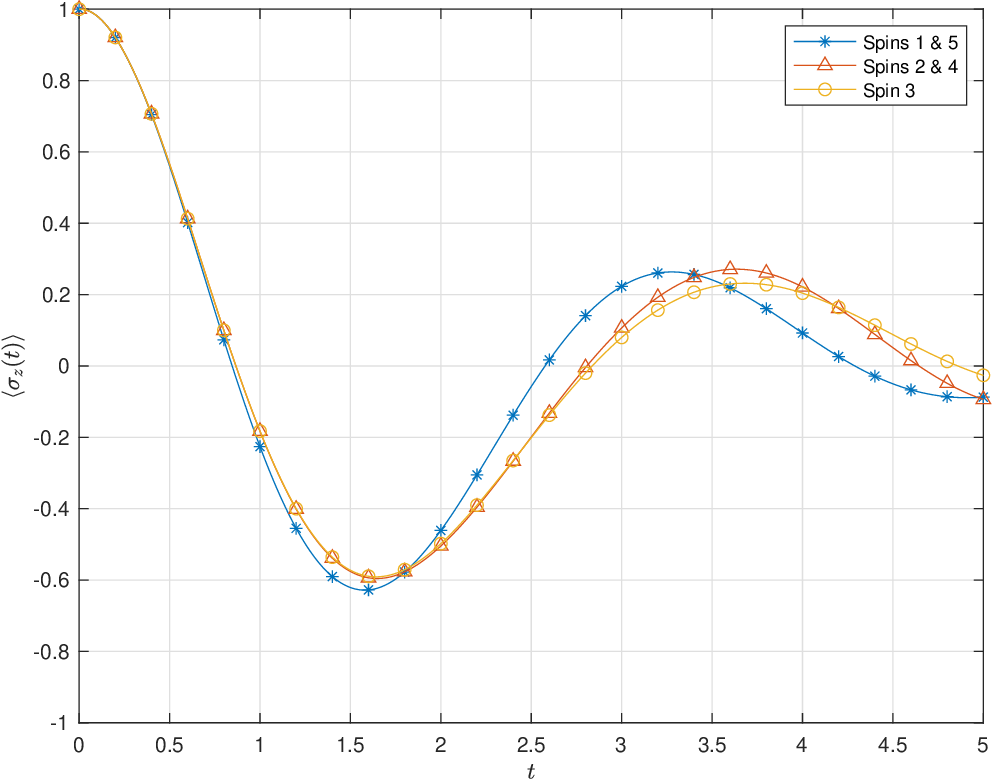}
         \caption{An Ising chain with 5 spins.}
         \label{fig_5_spins}
     \end{subfigure}
     \begin{subfigure}[b]{0.4\textwidth}
         \centering
         \includegraphics[width=\textwidth]{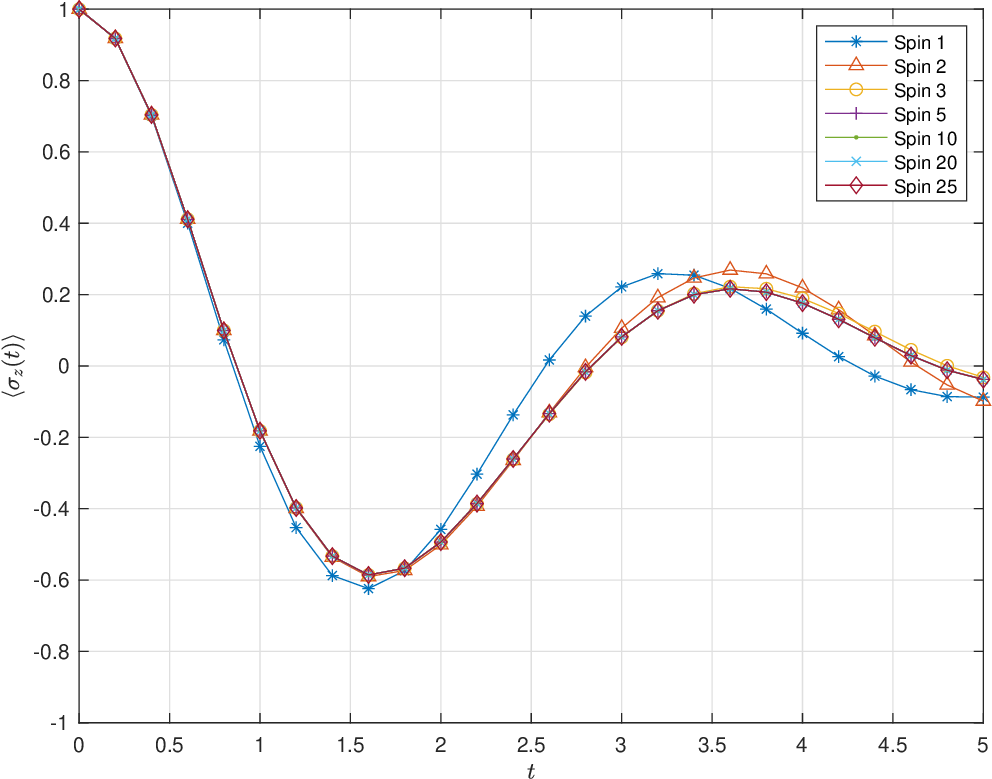}
         \caption{Observables of some spins in an Ising chain with 50 spins.}
         \label{fig_50_spins}
     \end{subfigure}
     \begin{subfigure}[b]{0.4\textwidth}
         \centering
         \includegraphics[width=\textwidth]{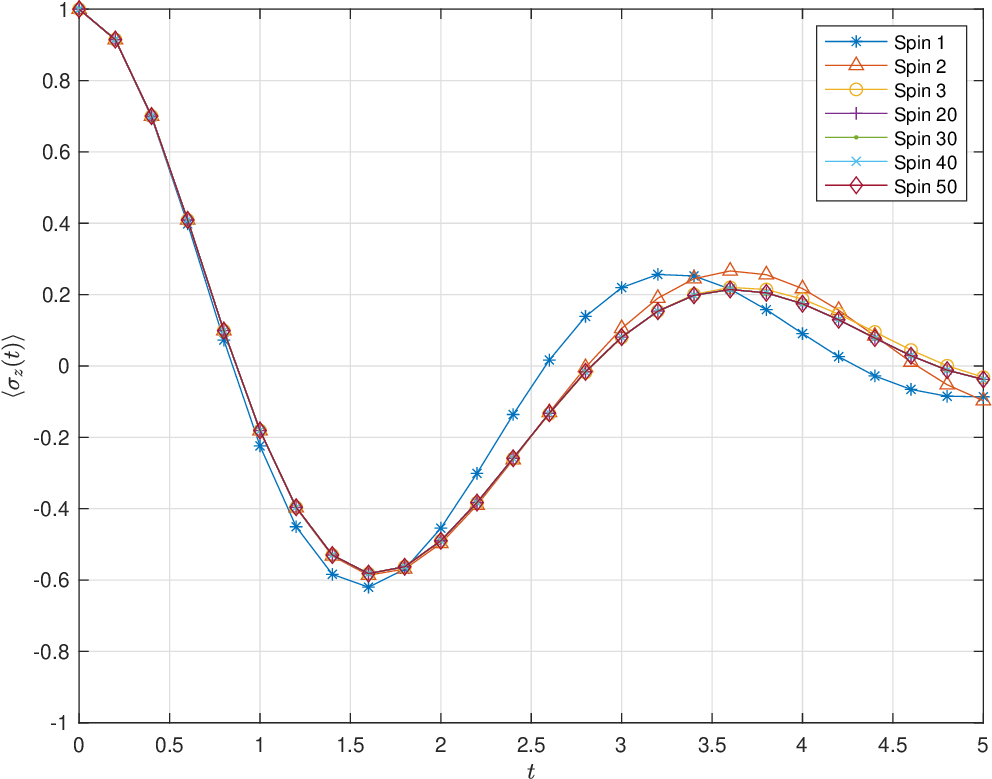}
         \caption{Observables of some spins in an Ising chain with 100 spins.}
         \label{fig_100_spins}
     \end{subfigure}
    \caption{Numerical Experiments for Ising chain with different length.}
    \label{fig_result_Ising_chain_with_different_length}
\end{figure}

Since all spins have the same parameters,
 the inchworm equation needs to be solved only once.
For longer spin chains, more computational cost is needed for the for the summation of full propagators.
But even so, according to our analysis in \Cref{sec_computational_cost},
 the summation only takes a small proportion of the computational time.
Our numerical results are presented in \cref{fig_result_Ising_chain_with_different_length}.
In general, the case of a single spin is clearly different from the interacting spin chains,
 while the three spin chains show very similar behaviors.
Due to the end effect,
 the first and the last spins have a slightly higher flipping frequency.
Between the third and the third last spin,
 the curves for all spins are indistinguishable in the plots,
 and in this example, the five-spin case can already well represent a long spin chain.


\section{Conclusion and Discussion}
\label{sec_conclusion}
We proposed a method to simulate an Ising chain coupled with harmonic baths.
The algorithm is derived by two steps: firstly, 
 the Dyson series decompose the system into spin-boson units 
 and the problem is also decomposed to a single spin problem;
 secondly, the inchworm algorithm is applied 
 to evaluate the evolution of spin-boson units with special ``crosses'' representing the spin-spin couplings.
The algorithm leads to the sum of diagrams.
 A special order for the summation based on distributive law is then proposed for faster evaluation of the sum,
 which accelerates the computation.
Under this special order for the summation,
 the most time consuming step is the computation for a single spin-boson unit.
The computational cost is then estimated by $\mathcal{O}(L^{\bar{M}+\bar{N}+2})$ 
where $L$ is the number of time steps 
and $\bar{M},\bar{N}$ are two truncation parameters for the series expansions.
Numerical experiments are carried out to validate our method.

While this paper focuses mainly on the Ising chain coupled with harmonic baths,
 similar idea can be migrated to more complicated interacting systems in a way similar to \cite{makri2018modular}.
Also, since our approach can be regarded as a perturbation theory,
 it is mainly applicable for short-time simulations.
Long-time simulations can be made possible by truncation of the memory kernel like the iterative QuAPI method.
These will be considered in our future works.

\bibliographystyle{abbrv}
\bibliography{myBib.bib}

\end{document}